\documentclass[%
 aip,
 pop,
 amsmath,amssymb,
 preprint,%
]{revtex4-1}

\usepackage{physics}
\usepackage{graphicx}
\usepackage{dcolumn}
\usepackage{bm}
\usepackage{xcolor}
\usepackage[caption=false]{subfig}
\usepackage{array}
\usepackage{float}
\usepackage{datetime}
\usepackage{cancel}
\usepackage{color}
\usepackage{comment}

\begin{document}

%\preprint{}

\title{Aligning Thermal and Current Quenches with a High Density Low-Z Injection}
\author{Jason Hamilton}
\email[]{jmhamilton@lanl.gov LA-UR-24-31891}
\author{Luis Chac\'on}
\author{Giannis Keramidas}
\author{Xian-Zhu Tang}
\email[]{xtang@lanl.gov}
\affiliation{Theoretical Division, Los Alamos National Laboratory, Los Alamos, NM 87545, USA}
\date{\today}

\begin{abstract}
The conventional approach for thermal quench mitigation in a tokamak
disruption is through a high-Z impurity injection that radiates away
the plasma's thermal energy before it reaches the wall.  The downside
is a robust Ohmic-to-runaway current conversion due to the radiatively
clamped low post-thermal-quench electron temperature.  An alternative
approach is to deploy a low-Z (either deuterium or hydrogen) injection
that aims to slow down the thermal quench, and ideally aligns it with
the current quench.  This approach has been investigated here via 3D MHD
simulations using the PIXIE3D code.  By boosting the hydrogen density,
a fusion-grade plasma is dilutionally cooled at approximately the
original pressure.  Energy loss to the wall is controlled by a Bohm
outflow condition at the boundary where the magnetic field intercepts
a thin plasma sheath at the wall, in addition to Bremsstrahlung bulk
losses.  Robust MHD instabilities proceed as usual, while the
collisionality of the plasma has been greatly increased and parallel
transport is now in the Braginskii regime.  The main conclusion of
this study is that the decreased transport loss along open field lines
due to a sufficient low-Z injection slows down the thermal quench rate
to the order of 20 ms, aligned with the current quench timescale for a
15~MA ITER plasma.
\end{abstract}


\maketitle

\section{Introduction}\label{sec1}

A major disruption in a tokamak is a sudden termination of the plasma discharge, which involves the removal of the plasma thermal energy and the magnetic energy associated with the plasma current.  A normal or
naturally occurring disruption has two distinct phases: a short thermal
quench (TQ) phase to rid of the plasma thermal energy and a relatively
longer current quench (CQ) phase to dissipate the plasma current.  For
a tokamak reactor like ITER, the thermal quench is projected to bring
down the plasma temperature from 10-20~keV by 2-3 orders of magnitude
over a time period of around one millisecond
(ms).~\cite{Hender:2007,nedospasov2008thermal} The current quench can
be much longer but the desired range, primarily for limiting the
electromagnetic force-loading in the blankets and vacuum vessel, is
around 50-150~ms for ITER.~\cite{lehnen2015disruptions} The reason why
the current quench can last a lot longer lies with two factors. The
first is induction in a plasma to resist the change in magnetic flux,
and the second is the radiative clamping of plasma temperature as the
consequence of a short ms-scale thermal quench. This second effect
comes about, in a naturally occurring disruption, because of the
intensive plasma-wall interaction as the chamber first wall and
divertor plates receive the plasma thermal pulse in the thermal
quench, which can generate plasma power flux 2-3 orders of magnitude
higher than that in steady-state operation. Since this transient
plasma power load can damage the first wall through localized melting, the
standard mitigation strategy for the thermal quench is high-Z impurity
injection, which aims to bring impurities into the plasma so it can be
radiatively cooled instead of onloading the plasma power to the
chamber
wall.~\cite{lehnen2015disruptions,shiraki2016thermal,commaux2016first}
In both un-mitigated and mitigated scenarios, a post-thermal quench
plasma is radiatively clamped to electron temperature as low as a few
eVs, as can be seen by balancing the Ohmic heating power with the
radiative cooling rate.~\cite{mcdevitt2022constraint}

A steep drop in $T_e$ implies much increased plasma resistivity as
$\eta \propto T_e^{-3/2},$ and so is the inductive parallel electric
field $\mathbf{E}\cdot\mathbf{B}=\eta \mathbf{j}\cdot\mathbf{B}.$ A
many orders of magnitude increase in inductive electric field as a
result of plasma temperature crashing from 10-20~keV to a few eVs, can
drive efficient Ohmic-to-runaway electron current conversion, through
the combination of runaway acceleration along the magnetic field
line~\cite{Dreicer59,connor-hastie-NF75,guo2017phase} and the
avalanche (thus exponential) growth of the runaway electrons due to
the knock-on collisions between primary runaways and the background
cold
electrons.~\cite{Jayakumar:1993,Rosenbluth97,hesslow-etal-prl-2017,McDevitt-2019}
A multiple mega-ampere current primarily made of a relativistic
electron beam encounters much reduced collisional drag, and hence
suffers a much more gradual decay even when the background electrons
are extremely cold.~\cite{guo2017phase} When the runaway electron beam
is guided by the magnetic field line onto the chamber wall, either in
the scrape-off process of an axisymmetric vertical displacement event
(VDE), or during a 3D MHD event that destroys the nested flux
surfaces, it poses a severe risk of localized melting and
drilling.~\cite{Hender:2007,artola2022non} Much recent work have been
on terminating the runaways safely on the first wall by strong
stochasticization of the magnetic field lines to spread the runaway
loads,~\cite{paz2021novel,reux2021demonstration,McDevitt-Tang-PRE-2023}
or on particulates in a standoff fashion.~\cite{Lively-etal-NF-2024}

To summarize the predicament we are in with regard to tokamak
disruption mitigation, it is the short duration of the thermal quench
that leads to radiative clamping of post-TQ plasma temperature, which
in turn enables robust and efficient Ohmic-to-runaway current
conversion. The extreme physics and engineering challenge that
disruption mitigation brings upon tokamak fusion is well-known and a
practical solution remains to be firmly
established.~\cite{eidietis-fst-2021} Returning to the root of the
problem, one may ask why the thermal quench of a 10-20~keV plasma is
so fast. The widely accepted culprit is 3D MHD activities that break
the nested flux surfaces, so the resulting 3D stochastic magnetic
field lines directly connect the fusion-temperature core plasma to the
chamber wall. The length of such open field lines, 3D by nature as
opposed to the 2D scrape-off layer, also known as magnetic connection
length $L_C,$ can produce very fast parallel transport if it is
sufficiently short.  The standard rule of thumb is that if $L_c$ is
comparable to or shorter than the mean-free-path of the fusion-grade
plasma $\lambda_{mfp},$ the collisionless parallel streaming would
produce an extremely fast thermal quench. The kinetic physics of this
parallel transport physics were previously described in
Ref.~\cite{zhang2023cooling,zhang2023electron,li2023staged,zhang2024collisionless}
for the cooling of both parallel and perpendicular electron
temperatures $(T_{e\parallel}, T_{e\perp}).$ Another interesting
finding from these previous studies, specifically discussed in
Ref.~\onlinecite{li2023staged}, is that even for a short magnetic connection
length, which corresponds to strong 3D MHD activities, the collisional
electron parallel transport and hence the thermal quench of $T_e$ are
comparatively far slower if impurity radiation is not a dominant
channel for plasma cooling.

This physics finding~\cite{li2023staged} motivates an alternative
approach in disruption mitigation in that if the plasma is placed into
the collisional regime before the thermal quench has taken place in
full effect, which means that a substantial fraction of the plasma
thermal energy has been transported out of the core, the thermal
quench would bypass the fast collisionless parallel cooling phase, and
directly land in the much slower collisional cooling regime.  In other
words, we want a way to increase the plasma collisionality but do not
cool the plasma temperature excessively.  A well-known approach is
dilutional cooling by massive injection of neutral hydrogen, for its
modest ionization energy to fully strip the atom.  For example, even
completely ignoring the energy loss due to ionization and radiation,
injection of hydrogen at 400 times the pre-disruption electron density
can dilutionally bring down the temperature by a factor of
approximately 400 times, notwithstanding the engineering challenge of delivering this
large load of hydrogen into the plasma and reprocessing the unburnt tritium afterwards.
For an initial core plasma at $T_e=10$~keV, that
implies a post-injection $T_e$ around 25~eV.  Exactly how long the TQ
would be as a result of collisional parallel transport will depend on
how strong the 3D MHD activities are triggered and sustained.  This is
the realm of extended MHD simulations that self-consistently account
for collisional transport and the magnetic reconnection physics
responsible for flux surface breakups as well as the evolution history
of $L_C$ in time and space.  The ideal outcome is the identification
of the physics and operational regime in which the thermal quench can
be prolonged enough that it overlaps with the current quench. This
offers the possibility, in the most ideal scenario, of a mitigated
disruption that avoids wall damage by both plasma thermal load and the
runaway electron beam. The former is due to the combination of longer
TQ duration and lower plasma power flux, as well as the lower plasma
temperature and hence modest ion energy impacting the wall that
reduces both wall damage and wall impurity production.  The thermal
load mitigation is further aided by the fact that, at much higher
plasma density, Bremsstrahlung radiation can contribute significantly
to plasma cooling and spread the power load over the entire chamber
wall.  The threat of a substantial runaway beam is mitigated by the fact that high
electron density is known to deplete the hot-tail runaway seed and
Ohmic heating of a hydrogenic plasma has the prospect of maintaining
the plasma at a warm enough temperature that the robust avalanche
growth of runaways is inhibited.

The primary objective of the current paper is to quantitatively assess the
feasibility of aligning the TQ and CQ in a mitigated disruption via
dilutional cooling of the plasma by massively raising the hydrogen
density.  This density would be orders of magnitude higher than the
famed Greenwald density, so violent MHD instabilities are to be
expected and a key aim of the extended MHD simulations is to quantify
how global magnetic chaos interacts with collisional plasma
transport. The density dependence of radiative cooling by
Bremsstrahlung radiation, in relation to stochastic field enhanced
collisional transport, is another important physics we hope to gain
insights into from extended MHD simulations with Bremsstrahlung
radiative cooling.

The rest of this paper is organized as follows.
Section~\ref{sec:simulation-setup} explains the set up of our extended
MHD simulations.  Section \ref{sec2} provides the MHD model used in
the simulations, including transport and boundary conditions.  The
specific simulation parameters of the PIXIE3D simulations will be
given in Section \ref{sec3}.  The results for various injection
densities will be shown in Section \ref{sec4}, while a contrasting
case without the sheath boundary is given in the appendix~\ref{sec44}.
Section \ref{sec5} will elaborate on these results and conclude that a
TQ can be significantly slowed down during a disruption to the same
timescale of the CQ, as well as ongoing and future work.

\section{Extended MHD simulation setup: physics considerations and simplifying assumptions\label{sec:simulation-setup}}

We have investigated the approach of TQ/CQ alignment via 3D MHD
simulations using the PIXIE3D code.  The PIXIE3D code and its solver
are described in Chac\'on (2004) and Chac\'on (2008)
respectively.\cite{chacon2004non,chacon2008optimal} Rather than
simulating the injection method itself (be it MGI or SPI or something
else), our simulations begin after the hydrogen density has been
boosted by a factor of, for example, 300, and the temperature has been
dilutionally cooled such that the original pressure is approximately
maintained (ionization and radiation will reduce the pressure from its
initial value, so to reach the same cooled temperature, the injected
density can be lower than that estimated by dilutional cooling
only). In our example of an ITER 15~MA equilibrium, a robust 1-1 kink
MHD instability is able to produce strong and global field line
stochasticity.  Several simulations are presented to highlight the
effects of the energy loss mechanisms that are available.  These loss
mechanisms consist of bulk losses due to radiation, and plasma
transport losses that eventually get out of the plasma through
conductive and convective boundary losses.  Bulk radiation loss is
provided by Bremsstrahlung, and we also assume an optically thin
plasma for simplicity. The subdominant or negligible contribution from line radiation
applies to a hydrogen plasma of temperature above a couple of (and
ideally ten's of) eVs, which is the regime we are aiming for to avoid excess
runaway production~\cite{mcdevitt2022constraint} for disruption mitigation.

As the collisionality of the plasma has been greatly increased by the
hydrogen injection, conduction throughout the plasma is
modelled by the near-equilibrium Braginskii (1965)
coefficients.\cite{braginskii1965transport} At low temperature, the
parallel thermal conduction (which has a $T^{5/2}$ scaling) along open
field lines is significantly reduced from the collisionless regime
value if no mitigation were to take place, as previously shown in
Ref.~\onlinecite{li2023staged}. In addition to conduction, there are
advective losses and collisionless conduction at the wall where a thin non-neutral sheath layer is
modelled between the computational domain and the first wall.  The
implementation of this sheath boundary condition, previously discussed
in Ref.~\onlinecite{tang-guo-nme-2017} using a kinetic sheath model from
Ref.~\onlinecite{tang-guo-pop-2016} and sheath energy transmission
coefficient data from Ref.~\onlinecite{tang2015sheath}, is similar to that
of Artola et al (2021)~\cite{artola2021simulations} and
Dekeyser et al (2021).\cite{dekeyser2021plasma} In our case, the Bohm
speed outflow condition uses the ideal Bohm speed of a collisionless
sheath plasma in a collisional plasma.~\cite{tang-guo-pop-2016L} The
non-ideal effects, namely the collisional modification of the heat
fluxes and temperature isotropization, have been found in
Refs.~\onlinecite{Li-etal-prl-2022,li2022transport} to modify the ideal Bohm
speed. But this more sophisticated treatment has not been implemented
in current MHD simulations.

To accommodate oblique incidence of magnetic field lines at the first
wall and divertor plates, our sheath boundary condition implementation
uses a critical grazing angle of $5^{\circ}$, below which there is
only cross-field diffusion into the wall.  This value falls in the
range determined experimentally by Matthews et al (1990), which
demonstrated that end losses along field lines that had grazing angles
below $5^{\circ}$ stopped following the anticipated cosine law, but
still had contributions from cross-field
diffusion.\cite{matthews1990investigation}

Our simulations use a perfectly conducting wall, which can impact the
disruption as both resistive wall tearing modes and vertical
displacement events (VDEs), as well as any other modes requiring a
non-ideal first wall, will not be present. Since an important aspect
of the TQ physics is the degree of field line stochasticity,
conveniently gauged by the magnetic connection length $L_c,$ a
conservative estimate on the TQ duration, which means a shorter
$\tau_{TQ},$ would be obtained in the strong MHD instability limit.
To this end, we have adopted a 15~MA H-mode ITER
plasma~\cite{Liu-etal-NF-2015} that is slightly modified by a
free-boundary Grad-Shafranov solver~\cite{Liu-etal-SIAM-JCC-2021} to
drive an even more violent (1,1) kink by further reducing the on-axis
safety factor $q_0.$ As we will show, this results in the
disappearance of an inversion radius and extremely strong global
magnetic stochasticity for a fast thermal quench in the weak
collisional limit.  In contrast, Strauss
(2021)~\cite{strauss2021thermal} has shown that resistive wall modes
(RWM) limit locked mode thermal quenches to 100 ms when the edge
plasma is collisional, which is significantly longer than the
resulting quench from the kink-unstable equilibrium used here. It can
be noted that their simulations used a very low beta equilibrium
($\beta = 0.008$ vs. $\beta = 0.028$ in the equilibrium used here) and
an unrealistically large perpendicular thermal conductivity.

With regard to the impact of the ideal wall boundary condition on
current quench, we note that for ITER, the vacuum vessel has a wall
time around 500~ms, and an acceptable CQ mitigation is supposed to
produce a CQ duration on the order of 100 ms or
less.~\cite{artola2024modelling} The VDEs on ITER or any other
reactor-type tokamaks are not going to be ideally unstable by design,
so the plasma current decay or CQ is mostly set by resistive decay if
a significant runaway population is avoided, which is the targeted
regime for our purpose. With a perfectly conducting wall, the plasma
column can shift but not scrape off, so there can be some shortening
of the CQ duration when non-ideal wall is included due to scrape-off.
This physics is not considered in current simulations and will be a
topic of a future study when the full torus version of the PIXIE3D
resistive-wall module~\cite{spinicci2023nonlinear} becomes available.

Our initial conditions assume that the hydrogen injection has already
been completed, thus no physics of either pellet ablation or
assimilation is included in the PIXIE3D simulations.  Since the large
injection density dwarfs any pre-injection density profile, the
post-injection density profile is assumed to be uniform throughout the
closed flux surface region, for lack of a better option.  The
resulting MHD instabilities and disruption may be affected by this
limiting density profile, although Commaux et al (2016) show
experimental evidence that the disruption following a pellet injection
vs a gas injection are not overtly sensitive to the difference in
density profiles.\cite{commaux2016first}

\section{Simulation models}\label{sec2}

This section will outline the physics model in the PIXIE3D simulations.
Section \ref{sec21} will describe the  MHD model, including transport and Bremsstrahlung.
Section \ref{sec22} will prescribe the boundary conditions for the sheath outflow losses.

\subsection{Extended MHD Model and Simplifications}\label{sec21}

For this study, PIXIE3D uses a single fluid, single temperature,
single ion species, quasi-neutral extended MHD model.  This model and
the fully implicit Jacobian-free Newton-Krylov solver are discussed in
detail in Chac\'on (2008).\cite{chacon2008optimal} Chac\'on et al
(2024) goes into detail about the numerical algorithm that allows us
to use realistic transport coefficients in regimes where the parallel
\& perpendicular thermal conductivity values are up to 7 orders of
magnitude apart.\cite{chacon2024a}

The standard MHD system of equations in SI units are:
\begin{equation}\label{cont0}
    \frac{\partial n}{\partial t} + \nabla \cdot \left(n \mathbf{U} - D \nabla n \right) = 0 \quad , 
\end{equation}
\begin{equation}\label{mom0}
    \frac{\partial \rho \mathbf{U}}{\partial t} + \nabla \cdot \left(\rho \mathbf{U} \mathbf{U} + \mathbf{\Pi}\right) = \frac{1}{\mu_0}(\nabla \times \mathbf{B})\times\mathbf{B} - \nabla  \left(2 n k_B T_e \right) \quad ,
\end{equation}
\begin{equation}\label{temp0}
    \frac{\partial T_e}{\partial t} + \mathbf{U}\cdot \nabla T_e + \left(\gamma -1\right) \left[ T_e \nabla \cdot \mathbf{U} + \frac{\nabla \cdot \mathbf{q} - Q}{2nk_B}\right] = 0 \quad ,
\end{equation}
\begin{equation}\label{faraday0}
    \frac{\partial \mathbf{B}}{\partial t} + \nabla \cdot \left(\mathbf{UB}-\mathbf{BU} \right) + \nabla \times \left(\frac{\eta(T_e)}{\mu_0} \, \nabla \times \mathbf{B}\right) = 0 \quad ,
\end{equation}
where $n$ is the ion density, $\rho=m n$ where $m$ is the mass, $\mathbf{U}$ is the plasma
velocity, $\mathbf{B}$ is the magnetic field, $T_e$ is the electron
temperature (assumed equal to the ion temperature), $\eta$ is the resistivity (given by the Spitzer model), $\mu_0 = 4\pi \times 10^{-7}$N/A$^{-2}$, $k_B = 1.38\times10^{-23}$J/K, $\mathbf{q}$ is a heat flux, $\mathbf{\Pi}$ is a deviatoric stress tensor, and $D$ is an \textit{ad hoc} particle diffusivity to allow
for cross-field diffusion.
The adiabatic index is $\gamma = 5/3$ (all
ions are fully ionized).

The set of equations which PIXIE3D solves is the dimensionless version of this system, which is:
\begin{equation}\label{cont}
    \frac{\partial n}{\partial t} + \nabla \cdot \left(n \mathbf{U} - D \nabla n \right) = 0 \quad , 
\end{equation}
\begin{equation}\label{mom}
    \frac{\partial n \mathbf{U}}{\partial t} + \nabla \cdot \left(n \mathbf{U} \mathbf{U} + \mathbf{\Pi}\right) = (\nabla \times \mathbf{B})\times\mathbf{B} - \nabla  \left(2 n T_e \right) \quad ,
\end{equation}
\begin{equation}\label{temp}
    \frac{\partial T_e}{\partial t} + \mathbf{U}\cdot \nabla T_e + \left(\gamma -1\right) \left[ T_e \nabla \cdot \mathbf{U} + \frac{\nabla \cdot \mathbf{q} - Q}{2n}\right] = 0 \quad ,
\end{equation}
\begin{equation}\label{faraday}
    \frac{\partial \mathbf{B}}{\partial t} + \nabla \cdot \left(\mathbf{UB}-\mathbf{BU} \right) + \nabla \times \left(\eta(T_e) \, \nabla \times \mathbf{B}\right) = 0 \quad ,
\end{equation}
which uses $m=1$, $\mu_0=1$, and $k_B=1$.
The ion density has been normalized to $10^{20}$m$^{-3}$, $\mathbf{U}$ has been normalized to the Alfv\'en speed $v_A = 1.18\times10^7$m/s, $\mathbf{B}$ has been normalized to $5.4$T, $T_e$ has been normalized to $723$ keV, and $\eta$ is now the inverse Lundquist
number.
All length scales are normalized to $L_0=2.18$m, and all time scales are normalized to $t_0 = L_0/v_A$.
All transport coefficients have been normalized by $L_0^2/t_0$.

The transport model is closed by using the collisional closure of
Braginskii (1965) for the heat flux $\mathbf{q}$ and a simple
hydrodynamic stress tensor for
$\mathbf{\Pi}$.\cite{braginskii1965transport} Since the assumed
mitigation conditions (discussed in Section \ref{sec3}) have reduced
the core plasma from $\sim20$keV to $\sim68$eV, such a closure is
appropriate, as the Knudsen number has been reduced from $K_n
\sim10^4$ to $K_n \sim10^{-3}$.  Although more accurate collisional
closures have been proposed more
recently,\cite{zhdanov2002transport,davies2021transport,hamilton2021formulation,hamilton2022plasma}
these results are very similar in the magnetized regime, so we retain
the Braginskii closure here for simplicity and its widespread use.  As
the magnetic field in a tokamak is dominated by the toroidal component
and the plasma is highly magnetized, the Nernst \& Ettingshausen
effects are ignored.  In addition, we opt for a simple hydrodynamic
stress tensor, leaving us with:
\begin{equation}\label{cond}
    \mathbf{q} = - \left(\chi_{e\parallel}  \mathbf{b} \mathbf{b} + \chi_{i\perp} \left(I -  \mathbf{b} \mathbf{b}\right)\right) \cdot \nabla T_e \quad ,
\end{equation}
\begin{equation}
    \mathbf{\Pi} = - \rho \nu_i \nabla \mathbf{U}  \quad,
\end{equation}
where parallel and perpendicular directions are defined by $ \mathbf{b}$, the magnetic field's unit vector.
We choose to ignore gyroviscous effects because we do not expect it to impact the dynamics. The main role of viscosity in our simulation is for regularization. We add artificial viscosity in the SOL (see Eq. \eqref{nu}) for this purpose which would dominate over any gyroviscous contributions.

The heating source $Q$ has contributions from Joule heating, viscous heating and a Bremsstrahlung sink term,
\begin{equation}\label{heat}
    Q = \eta J^2 - \mathbf{\Pi} : \nabla \mathbf{U} - P_{B} \quad .
\end{equation}
As the plasma is assumed to be optically thin, no absorption or any other radiation transport is considered apart from this energy sink term.
The plasma conditions remain above the Rayleigh-Jeans limit $T \gg h \omega$ and thus the classical expression for the power loss rate $P_B$ is used (Glasstone \& Lovberg, Controlled Thermonuclear Reactions 1960, Chapter 2)\cite{glasstone1960controlled},
\begin{equation}\label{bremNRL}
    P_B = (1.69\times 10^{-38} \,\, [\text{W/m\textsuperscript{3}}]) Z_i^3 n_i^2 \sqrt{T_{eV}} \quad ,
\end{equation}
where $T_{eV}$ is the electron temperature in units of eV.
Note that quasi-neutrality is assumed such that $n_e = Z n_i$.
We will combine the coefficient and all units into $P_{B0}$:
\begin{equation}\label{brem}
    P_B = P_{B0} Z_i^3 n_i^2 \sqrt{T_e} \quad ,
\end{equation}
where the charge state is assumed fixed at $Z_i = 1$.  A cutoff
temperature of 1 eV for this radiative cooling is implemented to avoid
cooling the wall plasma to zero before the disruption takes place.
This cutoff is only required because of the high initial density in
the pedestal region and the $n^2$ scaling of Bremsstrahlung.  A more
realistic density profile post-injection might not suffer from this
otherwise excessive cooling in this region.

The particle diffusion coefficient $D$ is isotropic but has a radial dependence,
\begin{equation}\label{diff}
    D = 10^{-5} \abs{1+9 r^{5}}, \quad r \,\in \,\left[0,1\right]
\end{equation}
such that the diffusion is an order of magnitude stronger in the
scrape-off-layer (SOL) than in the core, although the value of this
diffusivity is still very small and only allows a small amount of
additional momentum flux across the sheath boundary (described in the
next section). The interpretation of the logical radial coordinate $r$
in terms of physical ones is described in the
Appendix~\ref{app:iter-mesh}.  This diffusion into the wall prevents
large gradients from appearing there, which may cause numerical
issues.  A more physical boundary condition for the density would
include particle recycling from the sheath region, which may be
considered in future work.  As the computational domain extends beyond
the separatrix, $r=1$ corresponds to the first vacuum vessel wall, and
$r=0$ is the center of the computational domain at the initial
position of the magnetic axis.  The viscosity likewise has a radial
dependence,
\begin{equation}\label{nu}
    \nu_i = \nu_0 \abs{1+3 r^{5}}^4 \quad , \quad r \,\in \,\left[0,1\right]
\end{equation}
again such that there is higher viscosity in the SOL where the high
$\nabla T_e$ and low pressure combined with lower poloidal resolution
(we use an equally spaced fixed grid in $r$, $\theta$ and $\phi$) may
drive spurious oscillations from noise in this region.  This
additional viscosity damps these oscillations and prevents them from
propagating inwards.

The other transport coefficient dependencies are summarized as,
\begin{equation}\label{eta}
    \eta = \eta_0 T_e^{-\frac{3}{2}} \quad , \quad \chi_{i\perp} = \chi_{i\perp0} \frac{n^2}{\sqrt{T_{e}}B^2}\quad , \quad \chi_{e\parallel} = \chi_{e\parallel0} T_e^{\frac{5}{2}} \quad ,
\end{equation}
where all quantities are dimensionless.
Spitzer resistivity is used for $\eta$ based on the local temperature.
However, as the real Lundquist number for the collisional plasma can
still be too large to resolve with a reasonable simulation (thin
current layers near the $q=1$ surface are not resolved and must be
dissipated to avoid numerical issues), an artificially high $\eta$ (by
a factor of $\sim3$ globally) is used for some simulations.

Further care is taken near the vacuum vessel walls, where we floor the
value of $\chi_{\parallel}$ with that of $\chi_{\perp},$ considering that the low
temperature reduces the magnetization and the transport should be more
isotropic.  In addition, we set a vacuum ceiling of $\eta = 0.1$
(Lundquist number $S = 10$) to prevent large resistivities from
developing.

All other transport coefficients for viscosity ($\nu_0$), parallel \&
perpendicular thermal diffusivity
($\chi_{e\parallel0}$,$\chi_{i\perp0}$), and the Bremsstrahlung
radiation power ($P_{B0}$) are all the correct values based on
Braginskii (1965) for the local collisional plasma conditions used in
the simulations.  
Shown in Table \ref{tab:coefficients} are the coefficients used in the specific simulations presented in Section \ref{sec3}.
\begin{table}[]
    \centering
    \begin{tabular}{|c||c|c|c|c|c|c|c|c|c|}
    \hline
    injection & $n_o (m^{-3})$ & $T_0 (eV)$ & $\quad\eta \quad$ & $\nu_0$ & $\chi_{e\parallel0}$ & $\chi_{i\perp0}$ & $P_{B0}$ & TQ (ms) & CQ (ms) \\
    \hline
    $50 \times$    & $5\text{e}21$ & $410$ & $6\text{e-}6$ & $1.65\text{e-2}$ & $1.6$ & $3\text{e-}7$ & $4.03\text{e-}6$ & $\sim 2.5$ & $-$\\
    $300 \times$   & $3\text{e}22$ & $68$ & $6\text{e-}6$ & $1.9\text{e-}4$ & $1.8\text{e-}2$ & $7.7\text{e-}7$ & $1.45\text{e-}4$ & $\sim 20$ & $\sim 20$ \\
    $3000 \times$   & $3\text{e}23 $ & $6.8$ & $3\text{e-}5$ & $1.9\text{e-}4$ & $5.7\text{e-}5$ & $2.4\text{e-}6$ & $1.45\text{e-}2$ & $\sim 0.25$ & $-$\\
    \hline
    \end{tabular}
    \caption{Various state variables and coefficients for the 3 simulations presented in Section \ref{sec3}. All quantities are the initial values on the magnetic axis. TQ and CQ refer to the thermal and current quench durations observed in each simulated disruption. For the $50\times$ and $300\times$ cases, the duration is from the onset of stochasticity until the completion of the quench. For the $3000\times$ case, it is the entire duration of the simulation because no magnetic disruption occurred. The simulations ended when the TQ was over, so no CQ durations are given for $50\times$ or $3000\times$ because the current quench was slower.}
    \label{tab:coefficients}
\end{table}

\subsection{Sheath Boundary Model}\label{sec22}

The PIXIE3D simulations in the next section use the following boundary
conditions for the equations shown above.  At $r=0$ there is a
regularity condition for all
quantities.\cite{delzanno2008electrostatic} At $r=1$ there is a
homogeneous Neumann condition for the density and temperature in all
simulations with the sheath boundary, which shut down diffusive fluxes
at the boundary and allow the sheath boundary condition to account for
all boundary fluxes.  For the conduction-only case, the density and
temperature have non-homogeneous Dirichlet conditions set to the
initial wall values that allows for losses to the wall when material
builds up there even when there is no flow.  Since the boundary is
perfectly conducting, tangential magnetic field
components are held constant at the wall and the normal component
enforces the solenoidal constraint.

Since the PIXIE3D computational boundary is beyond the separatrix but
our equilibrium data is read from an EFIT file that only provides
flux-function quantities within the separatrix (except for the
poloidal flux itself), we make some simplifying assumptions about the
SOL.  The toroidal magnetic field is reconstructed assuming the
toroidal current is contained entirely within the separatrix, with the
field value outside determined by Stoke's theorem on Amp\`eres' law.
While toroidal currents outside this region may exist, we assume them to be small in equilibrium.
The temperature in the SOL is uniform and initialized at 1 eV.
The density is also uniform, set throughout the SOL to the post-injection density $n=C \times 10^{20}$m$^{-3}$, where $C$ is a different constant for each simulation.
The initial pressure in the SOL is therefore equal to $P_{SOL} = 16.02 C$ (units of N/m$^2$).

Finally, the velocity boundary condition allows for either a no-flow
Dirichlet condition, or an advective outflow based on an approximation
of a thin sheath layer residing between the computational boundary and
the wall, which we discuss in detail next.

As mentioned in Section \ref{sec1}, this sheath model is similar to
that of Artola et al (2021) and Dekeyser et al
(2021).\cite{artola2021simulations,dekeyser2021plasma} An advective
outflow is allowed along a magnetic field line that intercepts the
wall where there is assumed to be a thin sheath region.  Due to the
ambipolar electric field, ions are accelerated in the presheath toward
the sheath until they reach a critical velocity at the Bohm speed.
Ignoring the collisional effects, we have previously
shown~\cite{tang-guo-pop-2016L} that this outflow is at $U_B =
\sqrt{(T_e + 3T_i)/m_i} = 2\sqrt{T_e/m_i}$ using the upstream
temperatures (i.e., within the computational domain) and directed
parallel to the intercepting magnetic field in the direction going
into the wall, as long as the grazing angle between the magnetic field
and the wall surface is not too small. This is implemented in current
simulation with a threshold value $\theta_C$ for the grazing angle of
a few degrees, below which the Bohm outflow constraint is turned off.

From Eq. \eqref{temp}, one can see that the plasma energy loss due to
parallel transport along the magnetic field at the sheath entrance,
which is the boundary of our simulation, consists of two parts. One is
a conductive piece associate with the outflow at Bohm speed
($n_\alpha^{wall} T_\alpha^{wall} U_B$ with the species subscript
$\alpha=\{e,i\}$ denoting electron and ion populations). The other is
a conductive piece along the magnetic field at the sheath entrance,
$q_{\parallel,\alpha}^{wall},$ that is set by ambipolar collisionless
loss of charged particles since the the sheath is mostly
collisionless.~\cite{tang-guo-pop-2016,tang-guo-pop-2016L} Here the
superscript ``wall'' signifies that the quantities are taken at the
sheath entrance, which serves as the simulation boundary in lieu of
the actual wall.  It has long been the practice that the parallel
conduction can be parametrized in terms of multiples of the convective
energy flux, which is also called sheath energy transmission
coefficients ($\gamma_S$)~\cite{stangeby-book-2000}, in the form of
$q_\parallel = \gamma_S n_\alpha^{wall} T_\alpha^{wall} U_B.$ This
particular form has been found to hold by analytical theory for a
collisionless sheath in an otherwise collisional
plasma.~\cite{tang-guo-pop-2016,tang-guo-pop-2016L}

\begin{figure}
    \centering
    \includegraphics[scale=0.3]{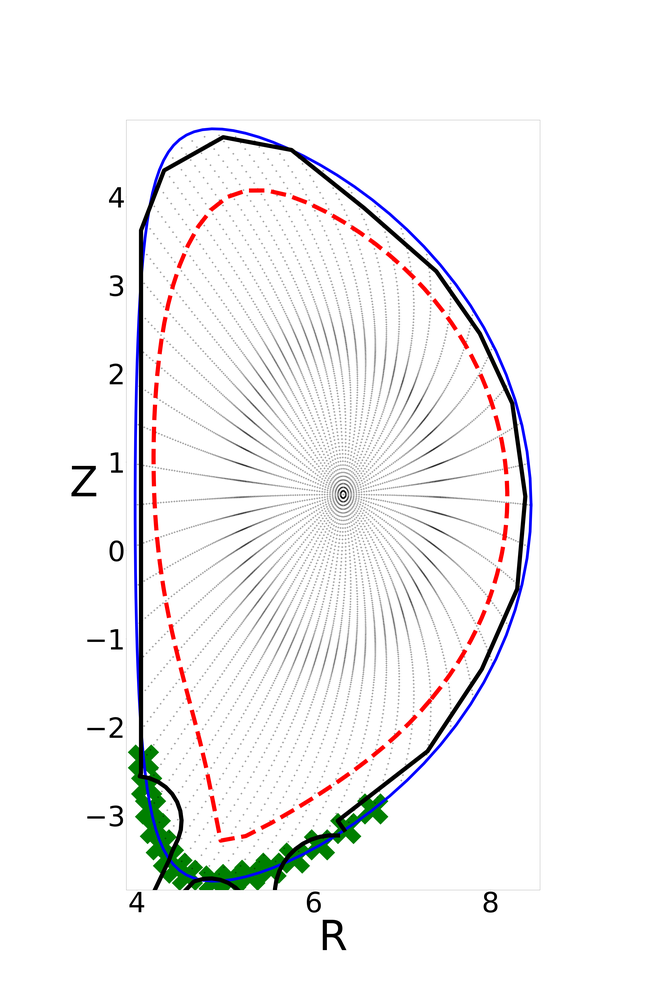}
    \caption{Geometry of the PIXIE3D simulations. The geometry and
      grid are toroidally symmetric with uniformly spaced grid points
      in the toroidal direction. The grid has dimensions
      $128\times64\times64$. Poloidal grid points are in gray, the equilibrium's
      separatrix is in red, the numerical boundary is in blue, and the
      actual ITER boundary is shown in black. Boundary grid points
      where the critical grazing angle is met is shown in green
      (described in Section \ref{sec22}). The coordinate map for these
      grid points is described in the Appendix.}
    \label{fig1}
\end{figure}

In current simulations, we implemented the sheath energy flux,
\begin{equation}\label{sheath2}
    \mathbf{q}_S = -\gamma_S P_{wall} U_B
    g_{\theta}\left(\abs{\frac{\arcsin{ \mathbf{b}\cdot
          \mathbf{n}}}{\theta_C}}\right) \mathbf{b} \, \,\,\text{sign}
    ( \mathbf{b} \cdot \mathbf{n})\quad ,
\end{equation}
where $ \mathbf{n}$ is the normal unit vector of the wall pointing
inwards, and the value $\gamma_S = 5$ is a rough approximation of the
kinetic value from Tang \& Guo (2015),\cite{tang2015sheath} owing to
PIXIE3D using a single temperature model requiring a combination of
the separate electron and ion effects.  The geometric factor
$g_{\theta}(x)$ contains the information about the grazing angle
$\theta_C$ as well as a transition function to smooth the sheath
energy flux between regions of finite and zero outflow,
\begin{equation}
    g_{\theta}(x) = \left(0.5 + 0.5 \tanh{\frac{10x - 5}{\sqrt{x(1-x)}}} \right) \quad ,
\end{equation}
providing the full Bohm outflow at normal incidence angles and zero
outflow at sufficiently sub-critical grazing angles.  Despite Artola
et al (2021) using a critical grazing angle of $2^\circ$, our value of
$5^\circ$ was chosen because the experimental result of Matthews et al
(1990) showed that the cosine law of parallel losses along field lines
breaks down at incidence angles between
$5^\circ-10^\circ$.\cite{artola2021simulations,matthews1990investigation}
Our results did show sensitivity to this value, with $\theta_C =
10^\circ$ providing essentially no advective outflow, and $\theta_C =
2^\circ$ being so lenient that the outflow was allowed around roughly
half of the boundary.
See
Figure~\ref{fig1} for a visualization of where the incidence angle
requirement was ultimately met for the chosen value of $\theta_C =
5^\circ$.

Particle recycling of ions and electrons being sent back into the
plasma from the sheath is not being modelled here, but will be the
subject of future work.

\section{Simulation Setup and the Underlying Physics Considerations}\label{sec3}

\subsection{Three distinct physics regimes under consideration for simulation studies\label{sec:physics-regimes}}

Thermal quench duration $\tau_{TQ}$ is set by the competition between
Ohmic heating and the two plasma cooling channels of plasma transport
loss and radiative cooling.  Let the ratio of post-injection density $n_e$
and pre-injection density $n_0$ be $C=n_e/n_0$. We can approximate the
post-injection plasma temperature, upon dilutional cooling alone, as
$T_e\sim T_0/C,$ with $T_0$ the pre-injection plasma temperature. The
reduction in plasma transport loss channel can be seen from the
post-injection electron parallel thermal conduction,
\begin{align}
\chi_{e\parallel} \sim \frac{n_e T_e \tau_e}{m_e} \propto \frac{C^{-5/2}}{\ln\Lambda}.
\end{align}
The Coulomb logarithm $\ln\Lambda$ has a weak dependence on $C$ so the
dominant scaling of $\chi_{e\parallel}(C)$ with respect to
high-density hydrogen injection and dilutional cooling is
$\chi_{e\parallel} \propto C^{-5/2}.$ The transport-induced cooling
time in a globally stochastic magnetic field is prolonged by a factor
of $C^{5/2}$ for the same magnetic field line connection length $L_c,$
\begin{align}
  \tau_{transport} \approx \frac{L_c^2}{\chi_{e\parallel}} \propto C^{5/2}.
\end{align}
There is the uncertainty that with lower temperature and higher
density, MHD instabilities might saturate to greater amplitude and thus
produce shorter $L_c.$ For a conservative assessment, we have chosen
an ITER 15~MA equilibrium that is modified to have a violent (1,1)
kink that has no inversion radius, the detail of which is given in
Sec.~\ref{sec:simulation-setup}.

Aligning the time scales of the TQ and CQ also implies that the magnetic energy dissipation
is directly involved in the TQ. The Ohmic heating power scales up with $C$ as a result of
dilutional cooling,
$P_{Ohmic} = \eta_\parallel j^2 \propto \left(\ln\Lambda\right) C^{3/2}.$
The ratio of transport loss rate and Ohmic heating rate, which is a good indicator
for the trend of $\tau_{TQ}/\tau_{CQ},$ scales as $C^4.$
For $C\sim 10^{2},$ we have a boost factor $C^4\sim 10^8$ approaching astronomical numbers.
The precise quantification is straightforward
for a fully ionized hydrogen plasma, which is the target plasma in our study. The parallel resistivity is
\begin{align}
  \eta_\parallel = 5.255 \times 10^{-5} \left(\ln\Lambda\right) T_e^{-3/2}\left(\textrm{eV}\right) \,\,\,\Omega \textrm{m}
\end{align}
where $T_e$ is in the unit of eV.
Expressing the current density $j$ in the unit of mega-ampere per squared meter (MA/m$^2$), one finds the
Ohmic heat power density
\begin{align}
P_{Ohmic} = 5.255 \times 10^{1} \left(\ln\Lambda\right) T_e^{-3/2}\left(\textrm{eV}\right) j^2\left(\textrm{MA/m}^2\right) \,\,\,\textrm{MW/m}^3.
\end{align}
In a post-injection plasma where $T_e\approx T_0/C,$ one finds
\begin{align}
P_{Ohmic} \approx P_{\eta 0} \left(\frac{j}{j_0}\right)^2
\frac{\ln\Lambda}{\ln\Lambda_0} C^{3/2}
\end{align}
with the pre-injection Ohmic heating rate %\lc{Here and below, use SI abbreviation for watt, W?}
\begin{align}
P_{\eta 0} = 5.255 \times 10^{1} \ln\Lambda_0
T_0^{-3/2}\left(\textrm{eV}\right) j_0^2\left(\textrm{MA/m}^2\right) \,\,\,\textrm{W/cm}^3.
\end{align}

The actual TQ and CQ durations, in the large $C$ limit for hydrogen injection, are subject to Bremsstrahlung
radiative cooling,
\begin{align}
  P_{Br} = 1.69 \times 10^{-32} n_e T_e^{1/2} \sum_Z \left(Z^2 n_Z\right) \,\,\,\textrm{W/cm}^3
\end{align}
where electron and ion number densities $n_{e,Z}$ are in the unit of
cm$^{-3},$ $Z$ is the charge state, and electron temperature $T_e$ is in
the unit of eV.
In a fully ionized hydrogen plasma,
\begin{align}
P_{Br} = 1.69 \times 10^{-32} n_e n_i T_e^{1/2} \,\,\,\textrm{W/cm}^3. 
\end{align}
With $n_i=n_e=n_0 C$ and $T_e \approx T_0/C,$ we find that
\begin{align}
P_{Br} \approx P_{Br0} C^{3/2}
\end{align}
where
\begin{align}
  P_{Br0}=1.69 \times 10^{-32} n_0^2 T_0^{1/2} \,\,\,\textrm{W/cm}^3
\end{align}
is the pre-injection Bremsstrahlung cooling rate.
To compare with the Ohmic heating rate, we write $n_0=\alpha \times 10^{14} \textrm{cm}^{-3},$ then
\begin{align}
  P_{Br0}=1.69 \times 10^{-4} \alpha^2 T_0^{1/2} \,\,\,\textrm{MW/m}^3.
\end{align}
Balancing $P_{Br0}$ with $P_{\eta 0},$ one finds a critical temperature $T_c$ for the pre-injection plasma beyond which
Bremsstrahlung radiation overpowers Ohmic heating,
\begin{align}
T_c =  \sqrt{3.1 \times 10^5 \ln\Lambda_0}\,\, \alpha j_0\left(\textrm{MA/m}^2\right).
\end{align}
For reactor plasmas, $\ln\Lambda_0 \approx 15-20, \alpha \approx 1,$
and $j_0 \approx 1-2$~MA/m$^2$ so $T_c \sim$~keV, which is less than a
pre-injection burning plasma at $T_e > 10$~keV.  In other words,
Bremsstrahlung radiation is usually greater than Ohmic heating before
high-density hydrogen injection ($P_{Br0} > P_{\eta 0}$), and because
of their same $C^{3/2}$ dependence, it remains so in the
post-injection plasma ($P_{Br} > P_{Ohmic}$), until further plasma cooling (TQ) and
the plasma current dissipation (CQ) can reverse the order, especially at the cooler plasma edge.

The competition of plasma transport cooling in a stochastic magnetic
field with a cooling time of $\tau_{transport} \propto C^{5/2},$ the
Bremsstrahlung cooling power $P_{Br} \approx P_{Br0} C^{3/2},$ and
Ohmic heating power $P_{Ohmic} \approx P_{\eta 0} C^{3/2},$ sets three
distinct physics regimes for high-density hydrogen injection.  In the
limit of modest $C$ injection, the Ohmic heating rate and
Bremsstrahlung radiative cooling rate remain sufficiently low that the
current quench time $\tau_{CQ}$ is much longer than the plasma cooling
time, which is dominated by $\tau_{transport}$ in the strong field
line stochasticity limit.  In this {\em first regime}, one recovers the usual
situation in which a rapid TQ is triggered once the MHD instabilities
saturate into sufficiently short $L_c$ such that a rapid TQ ensues
because of rapid parallel thermal conduction. The result is the usual experimental observation of
$\tau_{TQ}\ll \tau_{CQ}.$

When $C$ gets large, plasma transport loss is sufficiently slowed that
$\tau_{CQ}$ and $\tau_{transport}$ becomes comparable. It is important
to note that this particular or {\em second regime} is marked by a $C$ factor that is
not too large so the Bremsstrahlung radiation loss is comparable to
the plasma transport loss.  Successfully reaching this physics regime
would align the TQ and CQ by prolonging the $\tau_{TQ}$ to the time
scale of $\tau_{CQ}.$ The ideal target for optimal mitigation design
is to maintain a $T_e$ that stays above or close to the threshold
value for the parallel electric field to reach the runaway avalanche
threshold.

The {\em third regime} is reached by an even higher $C$ so Bremsstrahlung
cooling overwhelms plasma transport losses, even in the strong
stochastic magnetic field limit.  Since Bremsstrahlung cooling power
is always higher than Ohmic heating rate for a reactor-grade plasma,
$\tau_{TQ}$ is then dominated by Bremsstrahlung, which can be much
shorter than $\tau_{transport}$ in the strong stochastic magnetic
field limit for the large $C.$ The current quench, which is set by
Ohmic dissipation rate, now has $\tau_{CQ} \gg \tau_{TQ}$ again,
except that $\tau_{TQ}$ is driven by Bremsstrahlung radiation.  Since
the plasma energy is mostly radiated away, the power load on the wall
in the thermal quench phase is expected to be adequately
mitigated. The remaining concern is that if $T_e$ gets too low in the current
quench phase, runaway electron avalanche may be triggered for an efficient
Ohmic-to-runaway current conversion.

\subsection{Simulation setup for three prototypical scenarios\label{sec:simulation-setup2}}

\begin{figure}
    \centering
    \includegraphics[scale=0.5]{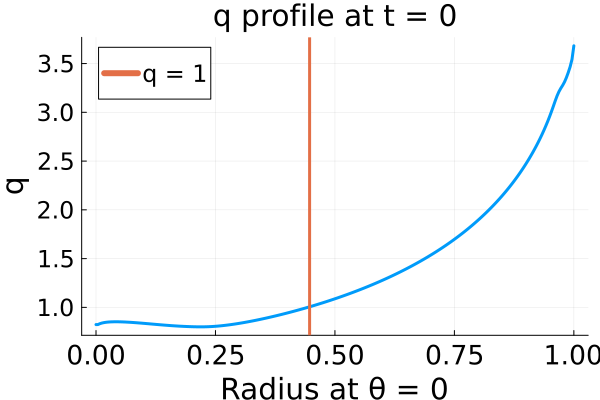}
    \caption{Initial q-profile for the equilibrium used in the simulations.
    This normalized radius is along $\theta =0$ from the geometric axis to the separatrix.
    The q-profile is extrapolated beyond the separatrix to the boundary.
    The vertical line shows the location where $q=1$ is crossed, which is where the internal kink mode appears which drives the disruption.}
    \label{q-profile}
\end{figure}

For all three prototypical scenarios that represent distinct physics
regimes introduced in Sec.~\ref{sec:physics-regimes}, we aim for a
conservative (low bound) for $\tau_{transport},$ which is realized in
the strong stochastic or short magnetic connection length $L_c$ limit.
To this end, we have modified a 15~MA H-mode ITER
equilibrium~\cite{Liu-etal-NF-2015} that is unstable to an $(n=1, m=1)$
internal kink, has a plasma beta of $\beta = 2.8\%$, field magnitude
$B_0 = 5.4$T, and minor radius $a = 2.18$m. The recomputed equilibrium
using a free-boundary Grad-Shafranov
solver~\cite{Liu-etal-SIAM-JCC-2021} retains the same plasma current
and the separatrix shape, but lower on-axis safety factor $q_0.$ The
q-profile of this modified equilibrium is shown in Figure
\ref{q-profile}, with $q=1$ surface around
$r/r_{sep}=0.45$.
We have
previously checked that there is no inversion radius for this unstable
equilibrium and the resulting global field line stochasticity is
strong.
Before the simulation begins, we have invoked the simplification that
the temperature has dilutionally cooled by the same factor as the
hydrogen density increase, such that the plasma pressure is unchanged.
In addition, the equilibrium is perturbed at $t=0$ with a small radial
velocity $V_r(r,\theta,t=0) = 10^{-4} \sin{\pi r}\cos{\theta}$.  The
fully implicit method described in Chac\'on (2008) is used with a
Jacobian-free Newton-Krylov solver.  In all cases, the timestep is
$\Delta t = 0.1 \tau_{A}$, where $\tau_{A}$ is the Alfv\'en
time.\cite{chacon2008optimal} The numerical coordinate map used in
these simulations is described in the Appendix~\ref{app:iter-mesh}.

For the three transport regimes of interest, we have chosen three representative injection
densities for simulation studies of the corresponding prototypical scenarios.
The target of choice for aligning thermal quench and current quench is
the second regime, for which we have chosen an injection density of  $300n_u$,
where $n_u$ is the
unmitigated density of $n_u = 10^{20} \text{m}^{-3}$.
For the first regime we have chosen 
$50n_u$, and for the third regime  $3000n_u.$   These density
increases will be denoted as \textit{$300\times$}, etc.
Next we go into some details about the simulation setups for each case.

For the prototypical simulation case for the first physics regime, one
might argue that the normal, unmitigated disruption would be a natural
choice. But such a simulation could not be performed with confidence
for multiple reasons. First, the $1\times$ case would be in a nearly
collisionless regime with $K_n = 10^{4}$, where Braginskii closure
does not apply and the collisionless thermal transport has been found
to differ from the free-streaming flux-limiting
form.~\cite{li2023staged} It remains a challenge to properly
incorporate this kinetic physics in an MHD code.  If we insist on  the
use of collisional transport equations in this regime, the thermal
transport anisotropy would be extreme at $\sim16$ orders of magnitude
with the normalized thermal conductivities $\chi_{e\parallel0} = 2.5\times10^{5}$ and $\chi_{i\perp0} =
1.6\times10^{-11}$.  Even with the novel fourth-order-accurate
transport model implemented in PIXIE3D that allows us to handle 7
orders of magnitude in anisotropy,\cite{chacon2024a} no known Eulerian
method is capable of 16 orders of magnitude without suffering
numerical pollution across field lines that degrades any physical
meaning to parallel losses.
Additionally, the normalized Spitzer resistivity value is inadequately small at
$\eta_0 \sim 10^{-11}$ to provide enough diffusion to handle
unresolved thin current sheets.  Thus to perform a simulation of the
unmitigated $1\times$ case we would have to artificially alter the transport
coefficients to such an extent that they would be devoid of any
physical semblance to the actual ITER conditions, rendering a study of
its transport losses not particularly meaningful.

We have picked the $50\times$ as the reference case for the first
physics regime.  By dilutionally cooling the plasma temperature by a
factor of 50, the plasma remains in low collisionality regime where
$K_n \sim 1$ in lieu of a true unmitigated case.  The normalized
transport coefficients, previously defined in Sec.~\ref{sec21}, are
now $\eta_0 = 2.7 \times 10^{-8}$, $\nu_0 = 1.65 \times10^{-2}$,
$\chi_{i\perp0} = 3\times10^{-7}$, $\chi_{e\parallel0} = 1.6$, and
$P_{B0} = 4.03\times10^{-6}$. As in all cases, Bremsstrahlung losses
will dominate over Joule heating for a net cooling effect in the
core. But at $50$x, the radiation loss is still far too weak to bring
a fast TQ.  Instead, the transport-dominated fast TQ will be set off
when the MHD instabilities saturate into global stochastic magnetic
field lines, at which point the rapid parallel thermal conduction
takes over.  It is interesting to note that a large parallel thermal
conduction flux must be accommodated at the boundary by a plasma
sheath.  This is primarily through the plasma density and temperature
at the sheath entrance, which directly enter the sheath power flux in
Eq.~(\ref{sheath2}) that scales $\gamma_s n_e v_{Bohm} T_e \propto
n_e^{1/2} T_e^{3/2}.$ In other words, the sheath can easily
accommodate a rising parallel thermal conductive flux by heating up
the boundary plasma temperature. This is the reason why a sheath
boundary condition for plasma outflow and sheath energy transmission
is critical to properly exhaust the conduction loss from the core.
To quantify this important point, we show the contrasting result
of a TQ simulation where the sheath boundary condition is replaced by
a conduction loss boundary condition in Appendix~\ref{sec44}.

For the opposite limit of high density injection in which
Bremsstrahlung would dominate the energy loss, we have picked an
injection density of $3000\times$ for the reference case. This is an
extremely collisional regime and the normalized transport coefficients
are $\eta_0 = 3 \times 10^{-5}$, $\chi_{i\perp0} = 2.4\times10^{-6}$,
$\chi_{e\parallel0} = 5.7\times 10^{-5}$, and $P_{B0} =
1.45\times10^{-2}$.  As the viscosity would be too small if left
unaltered, we retain the value $\nu_0 = 1.9 \times10^{-4}$ from the
$300\times$ case to prevent spurious oscillations from developing near
the boundary.  The Bremsstrahlung losses should quickly cool the core,
leading to a radiation-dominated rapid TQ.

For the second physics regime, which offers the plausibility of
aligning the time scales TQ and CQ by negotiating the relative
strength of Ohmic heating, transport loss, and Bremsstrahlung loss, we
have chosen the intermediate injection density of $300\times$. The
transport regime is collisional with a Knudsen number $K_n = 10^{-3}$.
The normalized transport coefficients are $\eta_0 = 10^{-6}$, $\nu_0 =
1.9 \times10^{-4}$, $\chi_{i\perp0} = 7.7\times10^{-7}$,
$\chi_{e\parallel0} = 1.8\times10^{-2}$, and $P_{B0} =
1.45\times10^{-4}$.

\section{Simulation Results \& Discussion}\label{sec4}

We present three PIXIE3D simulations with representative injection
densities described in section~\ref{sec:simulation-setup2} to contrast
TQ and CQ rates in different physics regimes. What an initial value
MHD simulation can offer beyond the scaling analysis in
Sec.~\ref{sec:physics-regimes} is (1) the confirmation of the analysis itself; (2) the self-consistent
evolution of $L_c$ when large scale MHD instabilities destroy the
nested flux surfaces to form globally stochastic magnetic field lines
that connect the core directly to the boundary; and (3) the radial
evolution of plasma temperature $T$ due to combined effects of
transport and radiation under Ohmic heating, so the variation of the
relative importance of Bremsstrahlung and transport losses in space
and time is self-consistently captured. 

\subsection{$300\times$ case}\label{sec41}

This PIXIE3D simulation starts after a massive hydrogen injection has
increased the ITER equilibrium density by a factor of 300, and the
temperature has dilutionally cooled by the same factor such that the
initial core temperature is now $\sim 68$eV.  The transport model
described above is used with the exception that the resistivity
coefficient $\eta_0$ needed to be increased by an \textit{ad-hoc}
factor (to $\eta_0 = 6 \times 10^{-6}$), which was found to be
the minimum value that is sufficient to diffuse under-resolved current
layers that formed in low pressure regions near the separatrix which
would otherwise lead to numerical issues.
Although this is larger than the nominal Spitzer value, we note that, in reality, a finite amount of anomalous resistivity may be present due to impurities, small scale turbulence, and other unresolved physics, providing uncertainty to the true resistivity.

\begin{figure}
    \centering
    \includegraphics[scale=0.2]{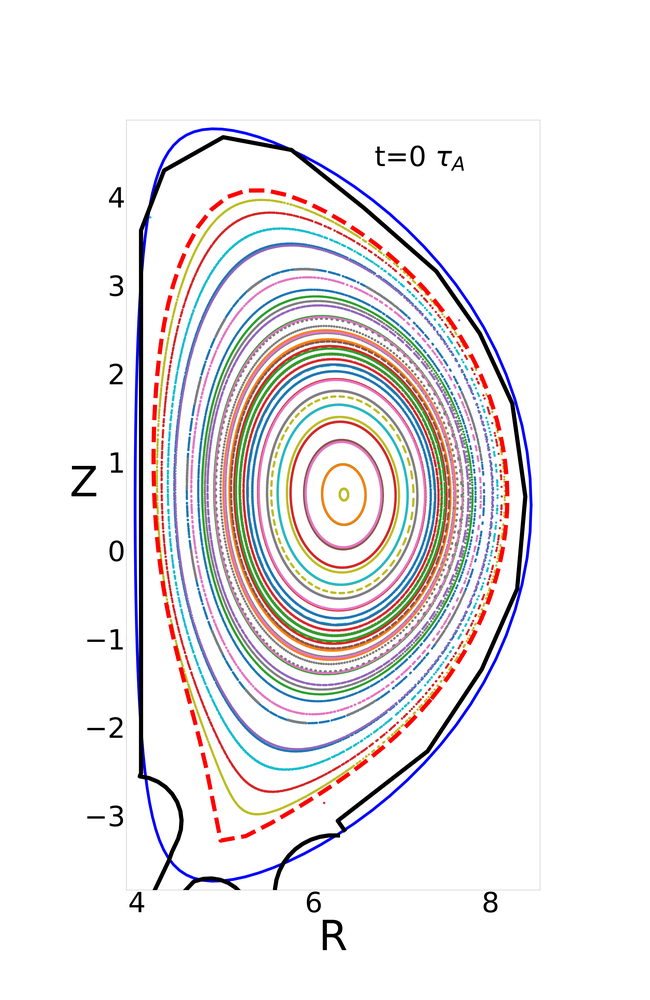}\label{fig6a}
    ~\includegraphics[scale=0.2]{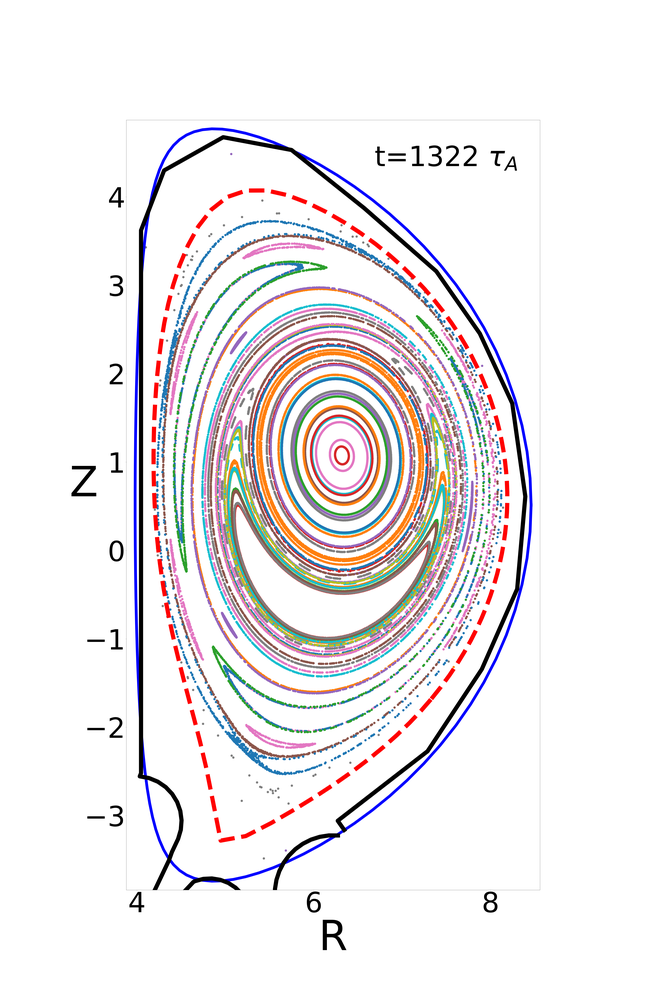}\label{fig6b}
    ~\includegraphics[scale=0.2]{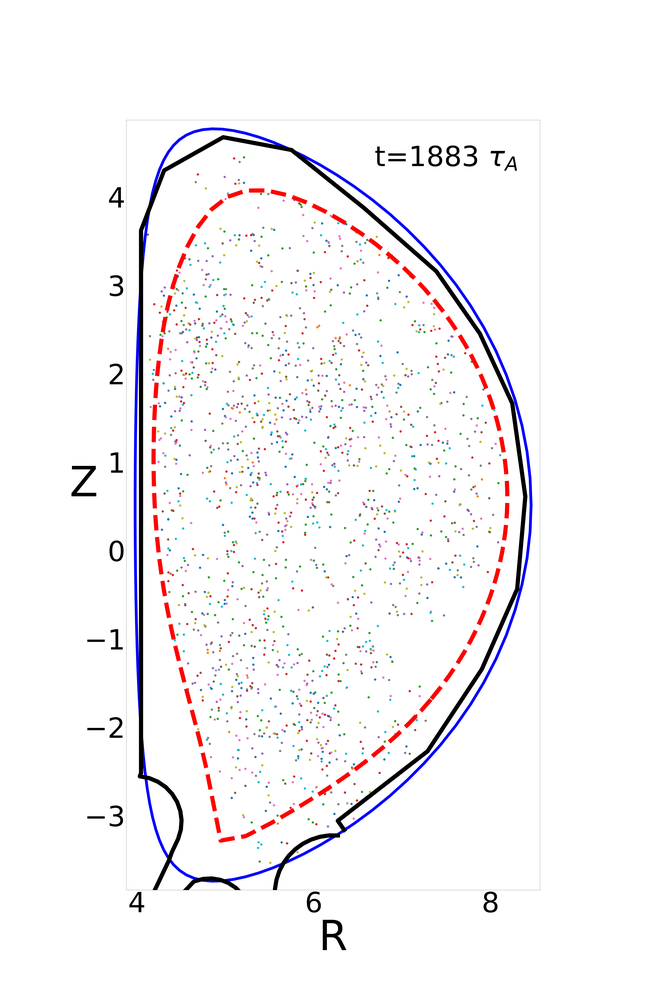}\label{fig6c}\\
    \includegraphics[scale=0.2]{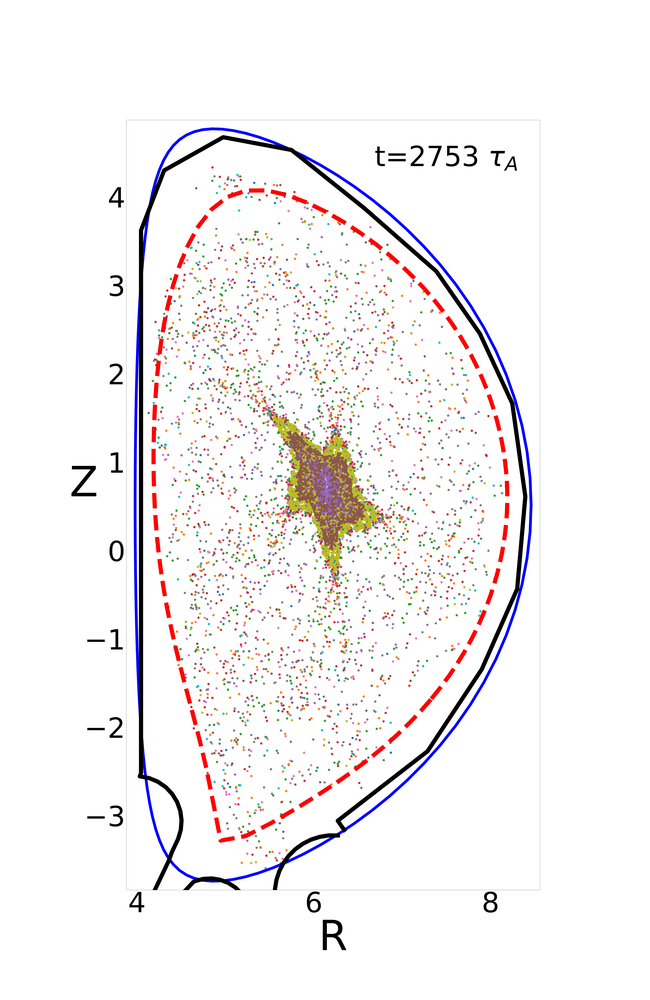}\label{fig6d}
    ~\includegraphics[scale=0.2]{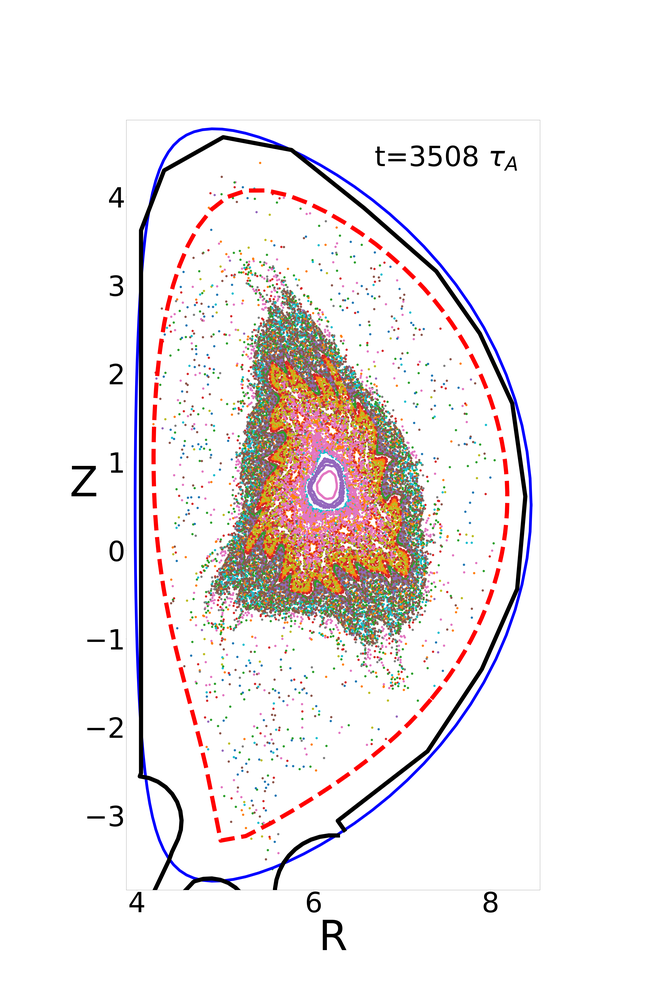}\label{fig6e}
    ~\includegraphics[scale=0.2]{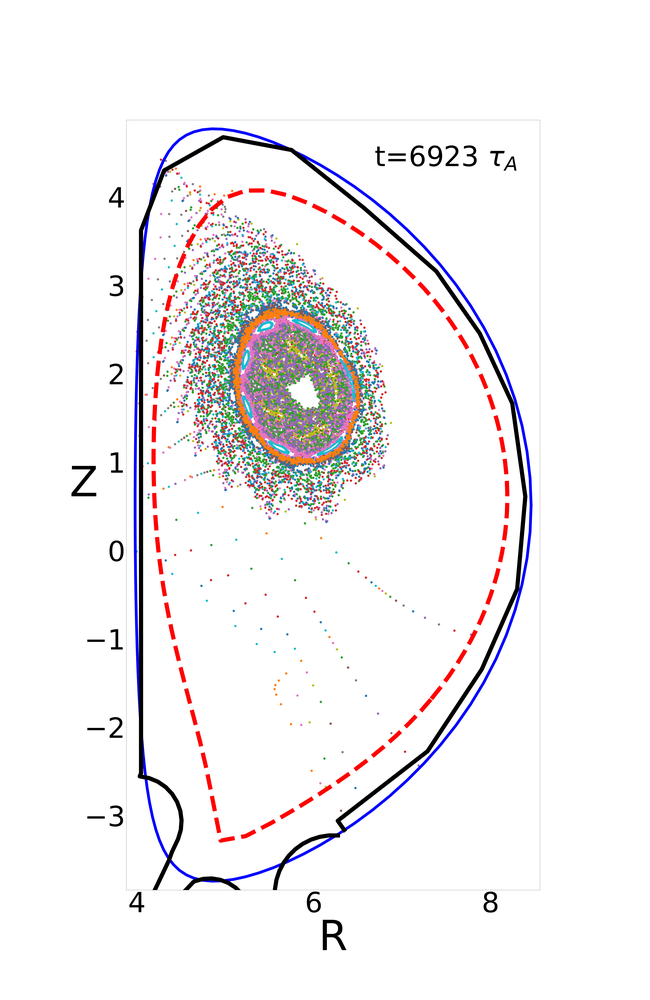}\label{fig6f}
    \caption{Poincare cross section plots of magnetic fields at
      various times throughout the $300\times$ simulation. The
      Alfv\'en time is $\tau_A = 3.2 \mu$s. The $(n=1, m=1)$ kink
      grows to large amplitude while higher m modes are also excited
      outside the $q=1$ surface all the way to the separatrix
      ($t=1332\tau_A$). Global stochasticity of the magnetic field
      lines is reached at $t=1883 \tau_A$. As the pressure drops, some
      core flux surfaces re-heal, but total plasma current and pressure continue
      to dissipate until they reach small values after approximately
      23 ms, when the simulation is stopped.}
    \label{fig6}
\end{figure}

\begin{figure}
    \centering
    \includegraphics[scale=0.2]{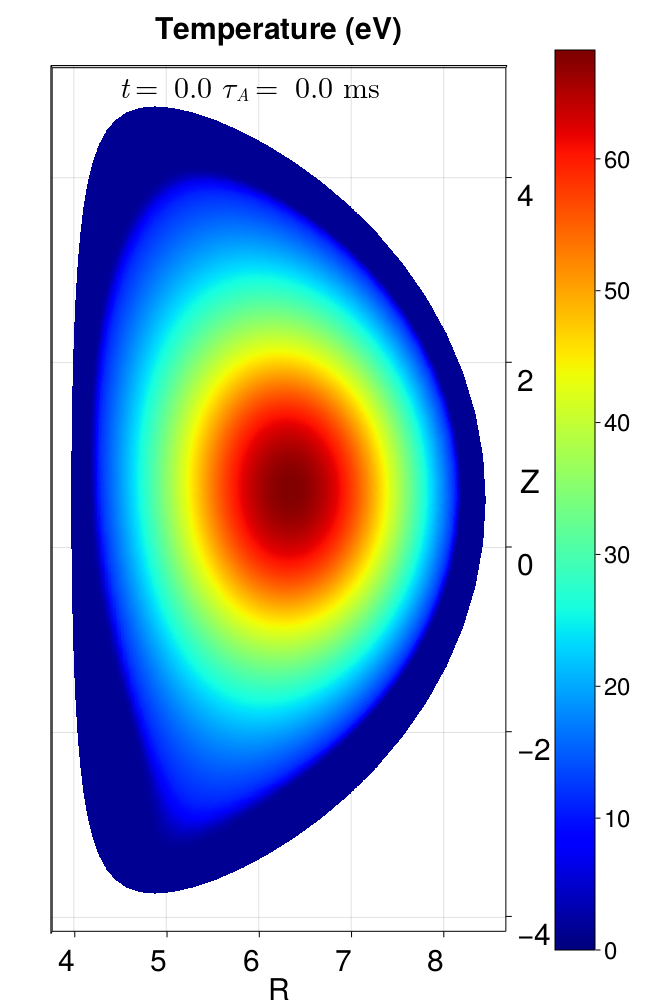}
    ~\includegraphics[scale=0.2]{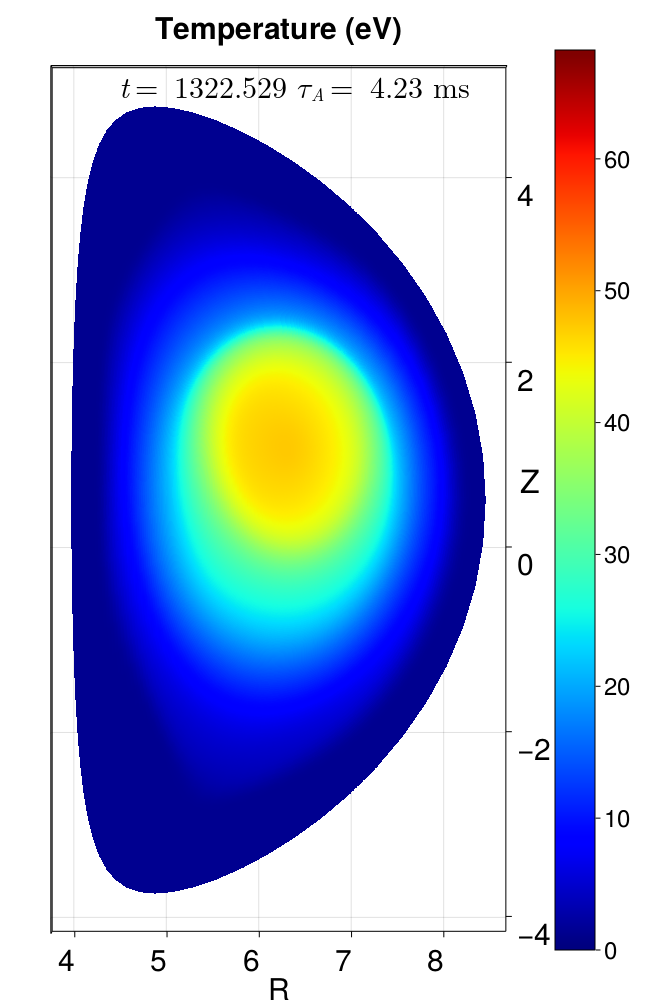}
    ~\includegraphics[scale=0.2]{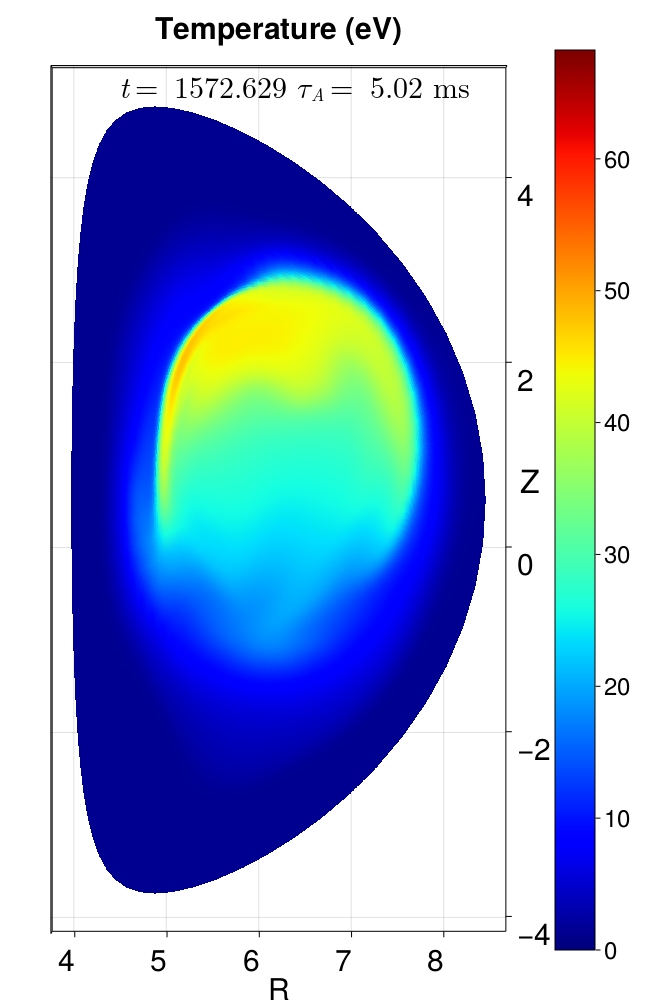}\\
    \includegraphics[scale=0.2]{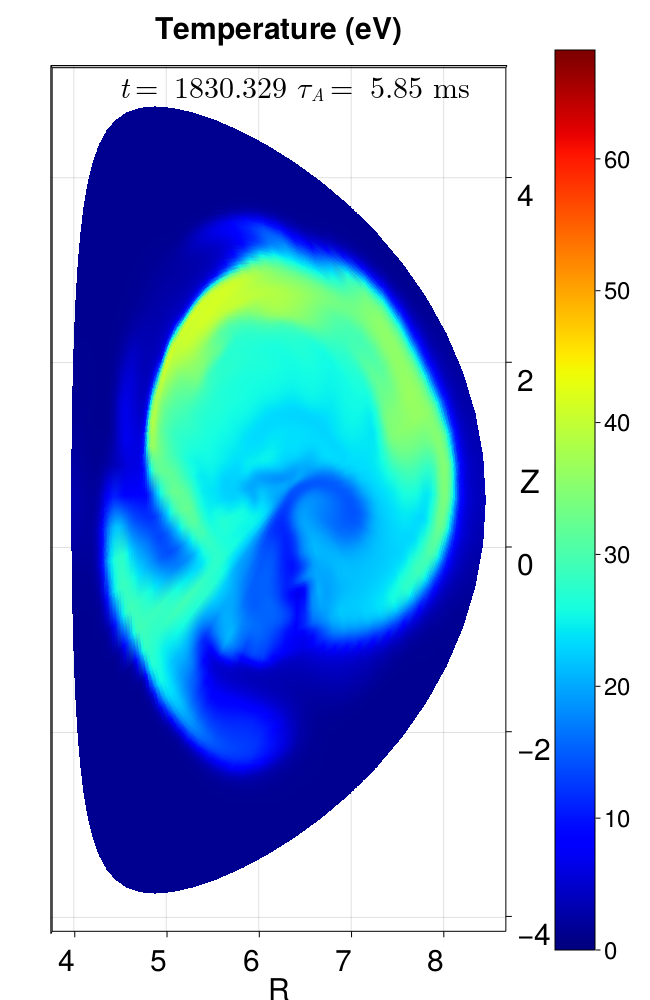}
    ~\includegraphics[scale=0.2]{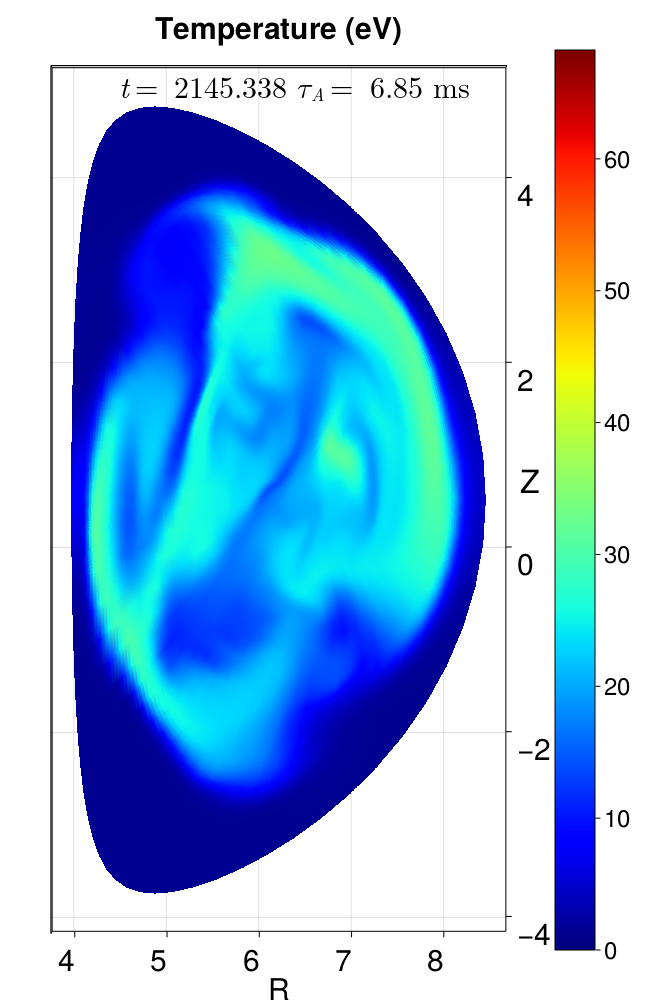}
    ~\includegraphics[scale=0.2]{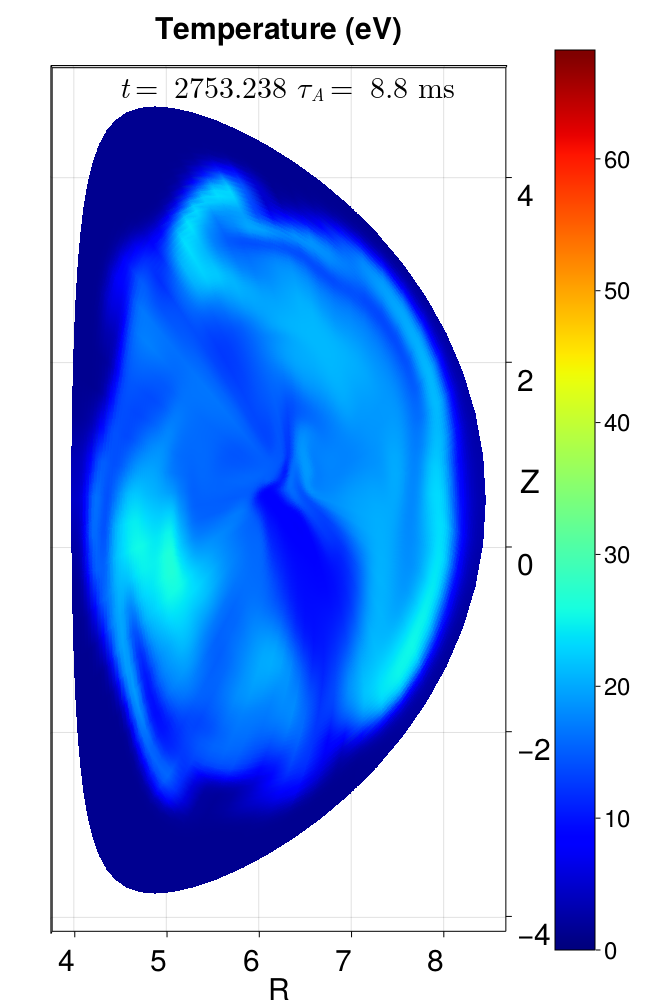}\\
    \includegraphics[scale=0.2]{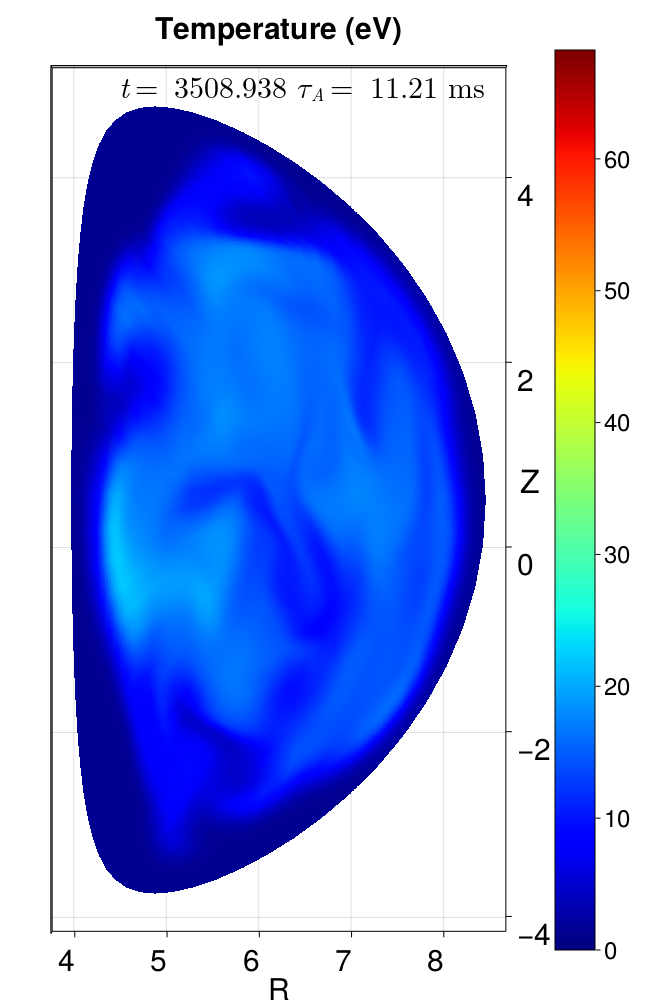}
    ~\includegraphics[scale=0.2]{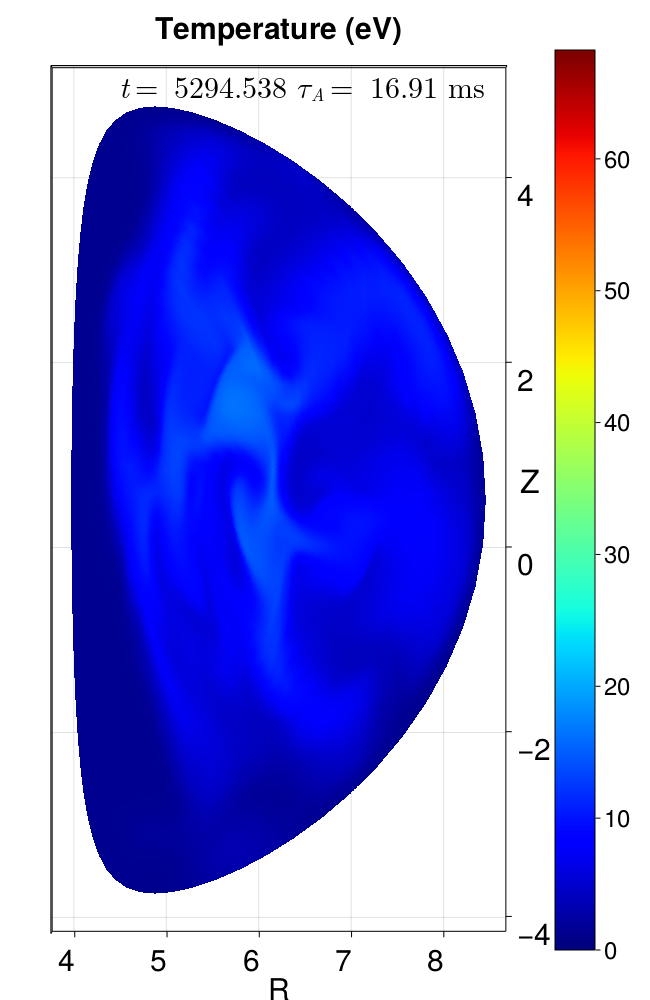}
    ~\includegraphics[scale=0.2]{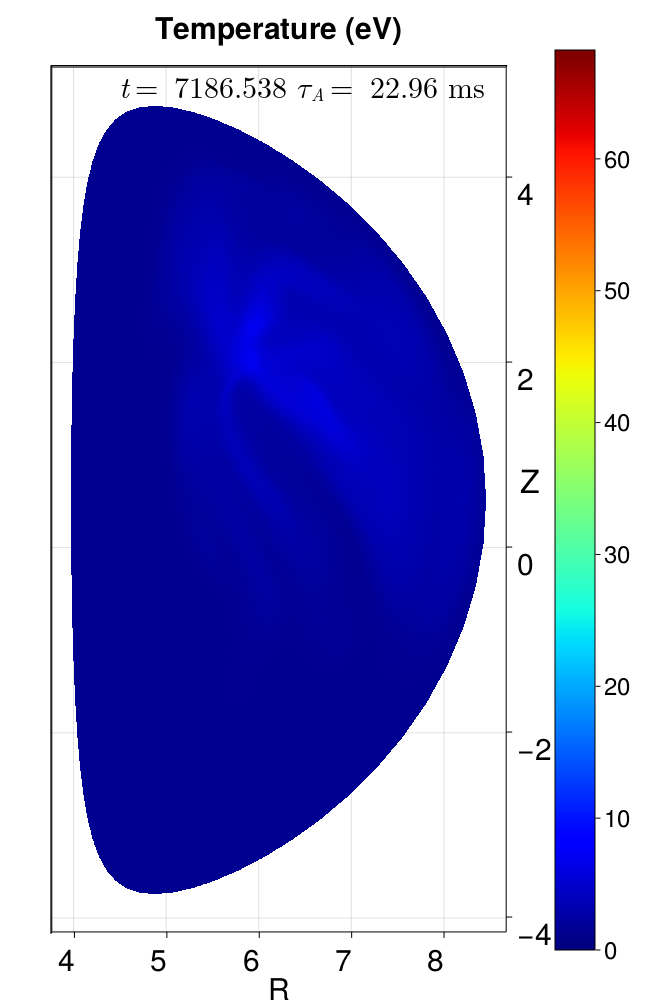}
    \caption{Temperature profiles at various times throughout the
      $300\times$ simulation, at $\phi=0$. The thermal energy has completely quenched after
      23 ms.}
    \label{fig7}
\end{figure}

The changes in magnetic field topology are shown in Figure \ref{fig6} for six 
time instances throughout the simulation.
Temperature profiles are shown in Figure \ref{fig7}.
The lack of an inversion radius for the (1,1) mode produces globally
stochastic magnetic field lines that allow parallel transport into the
walls, which aids the radial thermal conductive losses.  However, due
to the low value of $\chi_{e\parallel}$ (compared to the pre-injection
value of $\chi_{e\parallel0} = 2.5 \times 10^6$), this parallel
transport is sufficiently slow for a long TQ under Ohmic heating.

\begin{figure}
    \centering
    \subfloat{%
    \includegraphics[scale=0.35]{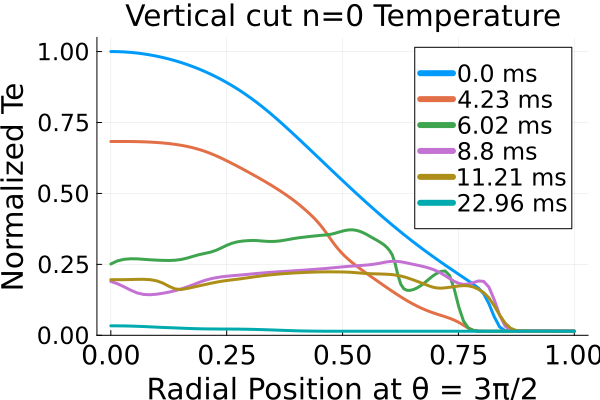}\label{fig8a}}\quad
    \subfloat{%
    \includegraphics[scale=0.35]{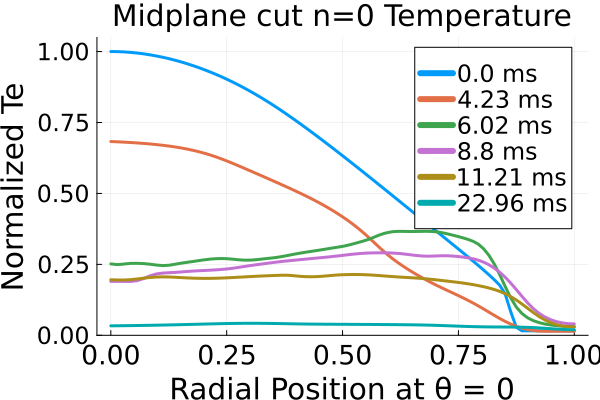}\label{fig8b}}
    \caption{Radial profile for the temperature from the magnetic axis
      (r=0) to the wall (r=1) for the $300\times$ simulation. Sheath
      losses near the divertor (left) result in lower temperature
      there than on the outboard side (right), but the entire system
      has a net cooling. The normalization value is 68~eV.}
    \label{fig8}
\end{figure}

The radial temperature profiles are shown in Figure \ref{fig8} at
various times for both a midplane radial cut ($\theta = 0$), and along
a vertical cut towards the divertor region ($\theta = 3\pi/2$).  As
the grazing angle threshold is only met near the divertor region (see
Figure \ref{fig1}), the energy losses are higher there.

\begin{figure}
    \centering
    \includegraphics[scale=0.35]{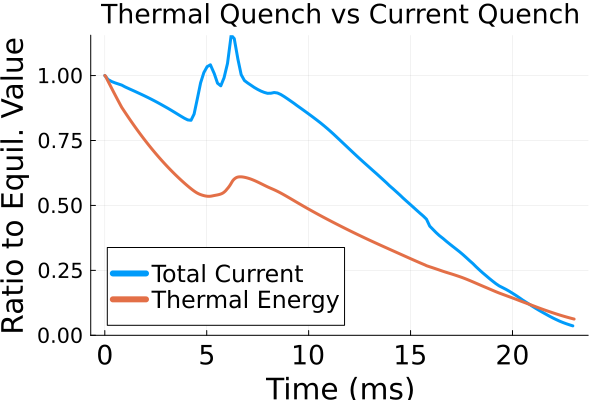}
    \caption{Time scales of the thermal \& current quenches are shown
      via the evolution of the total thermal energy and total current
      respectively, for the $300\times$ simulation. The current
      temporarily increases at the time of the breakup of the kink
      mode, see Figure \ref{fig6}. 
      The timescales for both the CQ and TQ are on the order of 20 ms
      post-disruption. The runaway current regime of low temperature
      and high current is completely avoided.}
    \label{fig9}
\end{figure}

Figure \ref{fig9} shows the time histories of thermal energy and total
current integrated across the entire domain and normalized to their
values at $t=0$.
Compared to the unmitigated scenario, the much faster CQ here can be
partly attributed to the higher resistivity, but also to strong field
line stochasticity that connects the parallel current channel from the
hotter core to the colder edge along open field lines.  As described
in Section \ref{sec:physics-regimes}, Bremsstrahlung dominates over Joule heating
and the plasma steadily loses energy before the 3D MHD instabilities
trigger a faster quench as a result of field line stochasticity.  As
it is well-known from tokamak disruption experiments and
high-resolution MHD simulations, a net plasma current bump can result
from current density profile relaxation in a globally stochastic
magnetic field.  The underlying physics has been previously invoked to
explain Taylor relaxation in low-beta plasmas.~\cite{} The gist of the
argument is that once the field lines become stochastic, pressure
relaxation or equilibration along the ergodic field lines can no
longer support an appreciable cross-field pressure gradient so the
perpendicular current density becomes negligibly small. This leads to
a plasma with mostly parallel current, $\mathbf{j} = \lambda
\mathbf{B}.$ Quasineutrality implies
\begin{align}
\nabla\cdot\mathbf{j} = \mathbf{B}\cdot\nabla \lambda = 0,
\end{align}
so $\lambda$ or $j_\parallel/B$ approaches a constant where the field
lines are stochastic.  A uniform $\lambda$ is the Taylor
state,~\cite{Taylor-prl-1986} but small amount of perpendicular current density
$\mathbf{j}_\perp$ can induce sizable modulation of $\lambda$ along
the field line via the Pfirsch-Schl\"{u}ter
effect.~\cite{Tang-Boozer-PoP-2004} More generally, to sustain the 3D
fields via MHD instability drive, some remnant radial gradient of the
averaged-$\lambda$ is usually retained in what is called a partially
relaxed plasma.~\cite{Tang-PRL-2007} In our specific case of a slow
TQ, radial pressure gradient will persist almost for the entire CQ
phase, so a partially relaxed, non-uniform $\lambda$ profile is to be
expected.

\begin{figure}
    \centering
    \includegraphics[scale=0.35]{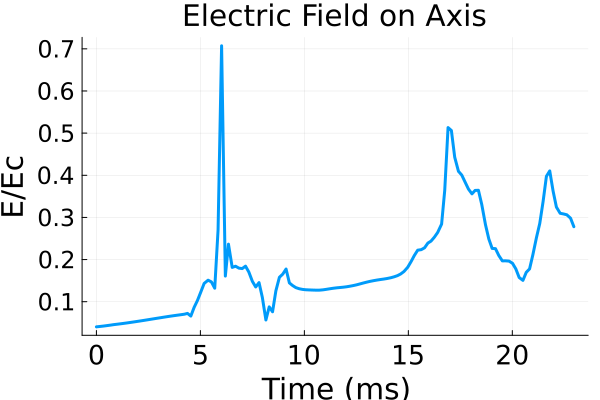}
    \caption{The parallel electric field at r=0 compared to the Connor-Hastie critical field value over time, for the $300\times$ simulation. This critical field $E_c$ determines when runaway electrons are generated but is not exceeded here.}
    \label{figEfield300}
\end{figure}

Interestingly, our simulation not only shows a significant plasma
current spike, but also an obvious thermal energy bump.  Following the
onset of stochasticity (around 5 ms), thermal quench is accomplished
over a time duration of 20~ms, during which the magnetic field lines
remain largely stochastic except for the very core region.  This is an
order of magnitude longer than the unmitigated scenario (see
section~\ref{sec42}), usually estimated to be around 1
ms~\cite{nedospasov2008thermal}, but can be much shorter in actual
experiments~\cite{riccardo2005timescale}, and was recently found to be
a function of $L_c$ in parallel conduction dominated
regime.\cite{li2023staged} Most importantly, the timescales of the TQ
and CQ are identical in this case, meaning that the conditions for runaway current
generation (low temperature, high current) are also
mitigated. 
Figure \ref{figEfield300} shows the on-axis parallel electric field magnitude, $|\mathbf{E\cdot \hat{b}}|$, compared to the Connor-Hastie critical field value $E_c,$~\cite{connor-hastie-NF75}
\begin{equation*}
    E_c = \frac{n_e e^3 \ln{\Lambda}}{4\pi \epsilon_0^2 m_e c^2} \, ,
\end{equation*}
which is a conservative value for the true critical field according to Stahl et al (2015).\cite{Stahl-PRL-2015}
The electric field does not exceed this critical value so runaway electrons are not expected to be generated.
The current spike seen in Figure \ref{fig9} corresponds to a sharp jump in the parallel electric field shortly after 5 ms.

\subsection{$50\times$ case}\label{sec42}
Reducing the injection density from $300\times$ to $50\times,$ the
parallel thermal conductivity is increased to $\chi_{e\parallel0} =
1.6$, while the resistivity is now $\eta = 10^{-8}$.  However, the
resistivity must again be artificially raised once the disruption (onset of stochasticity) occurs to $\eta_0 = 6 \times 10^{-6}$ to avoid
numerical issues.

\begin{figure}
    \centering
    \includegraphics[scale=0.35]{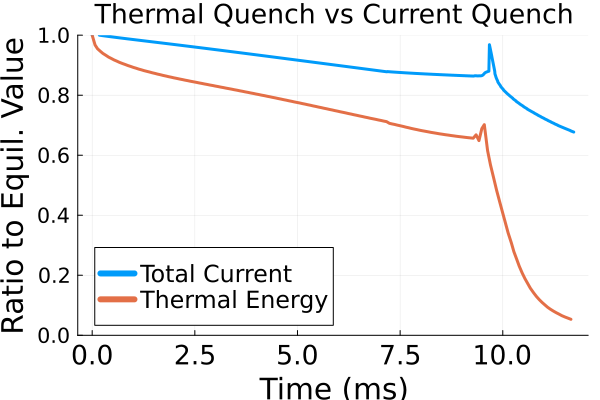}
    \caption{Time scales of the thermal \& current quenches are shown
      via the evolution of the total thermal energy and total current
      respectively, for the $50\times$ simulation. Post-disruption
      parallel conduction losses lead to a rapid TQ, but the current
      decays at a much slower rate.}
    \label{fig9b}
\end{figure}

\begin{figure}
    \centering
    \includegraphics[scale=0.25]{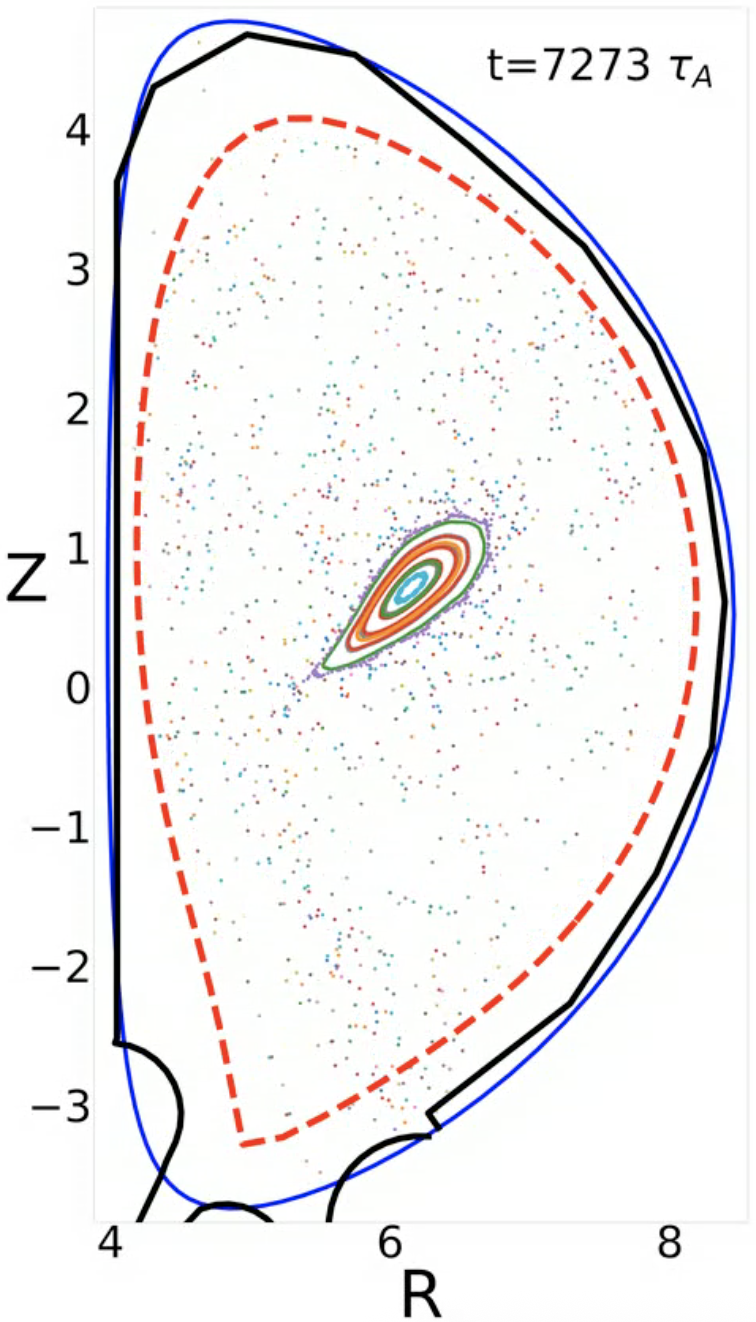}
    ~\includegraphics[scale=0.25]{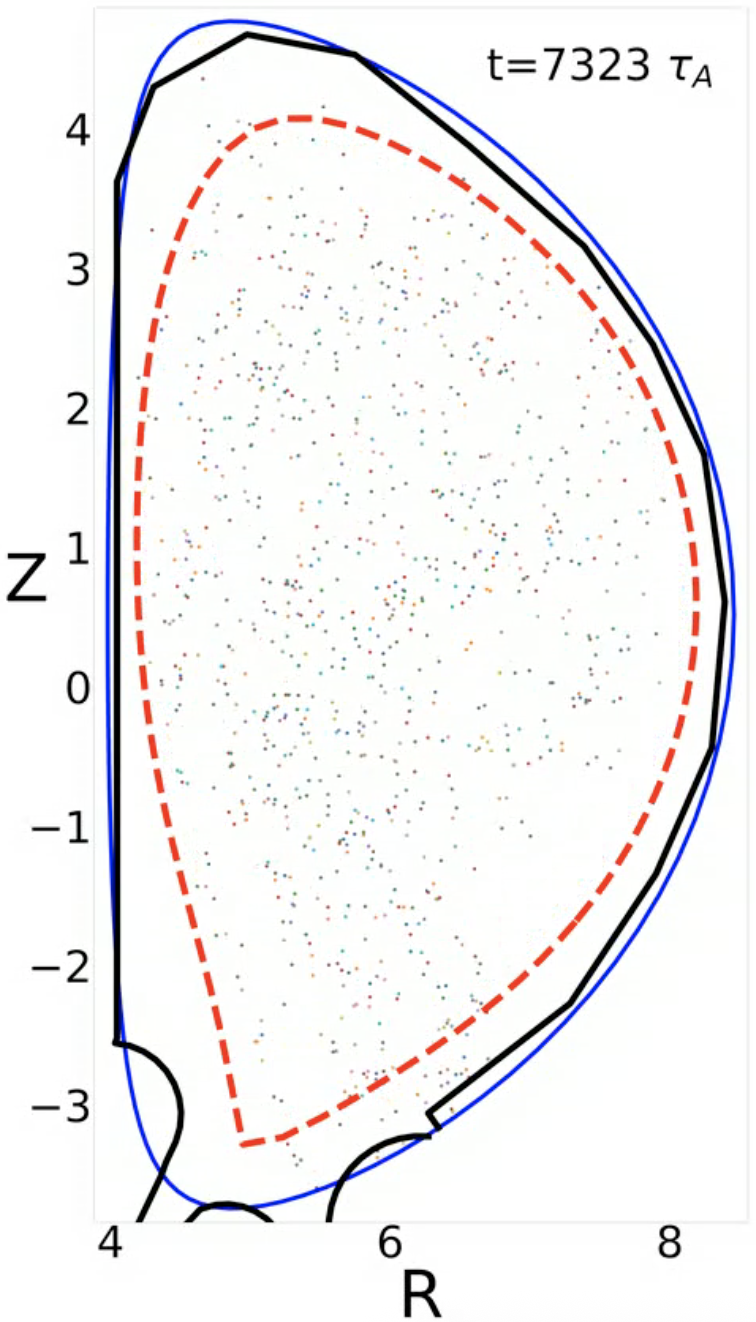}
    ~\includegraphics[scale=0.25]{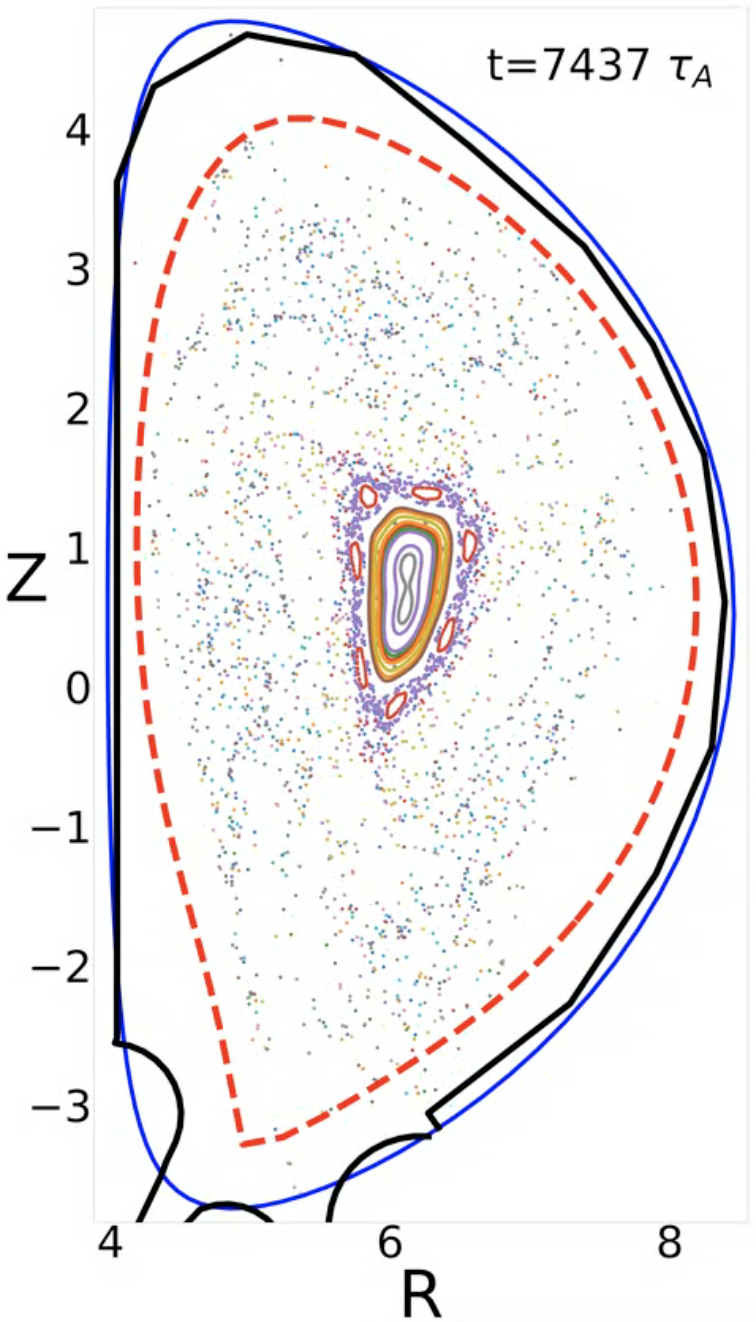}
    ~\includegraphics[scale=0.25]{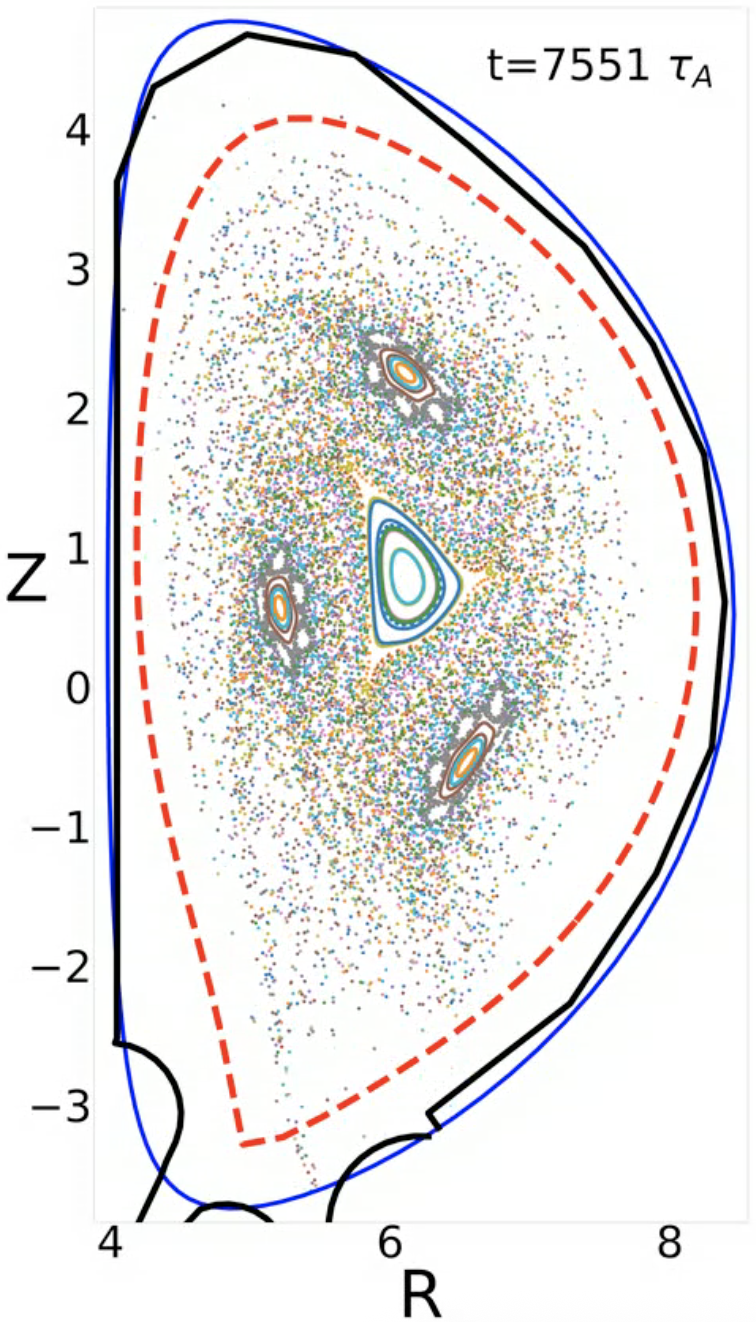}
    \caption{Poincare cross section plots of the magnetic field at $\phi=0$ immediately before (7273 $\tau_A$), during (7323 $\tau_A$), and after (7551 $\tau_A$) the current spike of the $50\times$ simulation. Magnetic reconnection events occur during this current spike. The TQ begins with the onset of stochasticity around the same time as the current spike. The
      Alfv\'en time is $\tau_A = 1.3 \mu$s.}
    \label{fig50xpoincare}
\end{figure}

\begin{figure}
    \centering
    \includegraphics[scale=0.35]{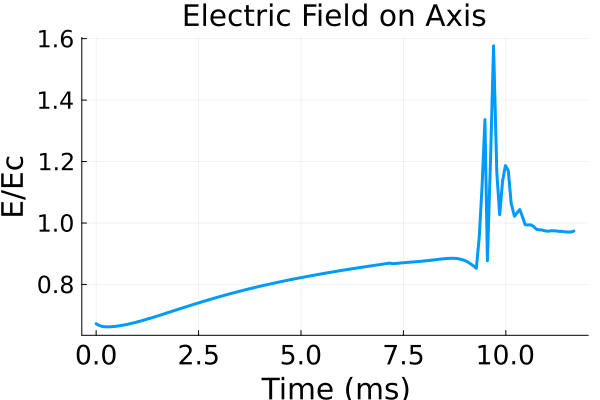}
    \caption{The parallel electric field at r=0 compared to the Connor-Hastie critical field value over time, for the $50\times$ simulation. This lower injection density is at more risk for runaway generation compared to  the $300\times$ case in Figure \ref{figEfield300}.}
    \label{figEfield50}
\end{figure}

The higher parallel conductivity in this case has very little effect
on the system before the disruption.
The total thermal energy and current are shown in
Figure \ref{fig9b}.
The evolution of the temperature profiles and magnetic topology are
similar to the previous example in Section \ref{sec41}.
Poincare section plots during the current spike are shown in Figure \ref{fig50xpoincare}.
A notable
difference is that with higher $\chi_{e\parallel}$, thermal energy
gets to the wall faster following the onset of stochasticity, which
leads to a faster drop in temperature.
The on-axis electric field compared to the Connor-Hastie field is shown in Figure \ref{figEfield50}
and shows that the critical field (or a conservative estimate for the critical field\cite{Stahl-PRL-2015}) is exceeded during the current spike, which begins around $9.5$ms.

As for the CQ, the lower resistivity leads to a slower decay of
current, despite the increased parallel conductivity.  From the
inbalance in these loss rates, it is evident that the CQ is not
aligned with the faster TQ at these conditions.

\subsection{$3000\times$ case}\label{sec43}
\begin{figure}
    \centering
    \includegraphics[scale=0.35]{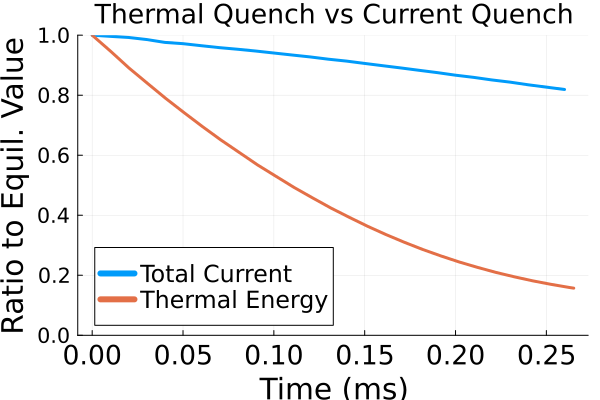}
    \caption{Time scales of the thermal \& current quenches are shown
      via the evolution of the total thermal energy and total current
      respectively, for the $3000\times$ simulation. A rapid TQ
      immediately follows the massive hydrogen injection, owing to
      large Bremsstrahlung losses.}
    \label{fig10}
\end{figure}

\begin{figure}
    \centering
    \includegraphics[scale=0.3]{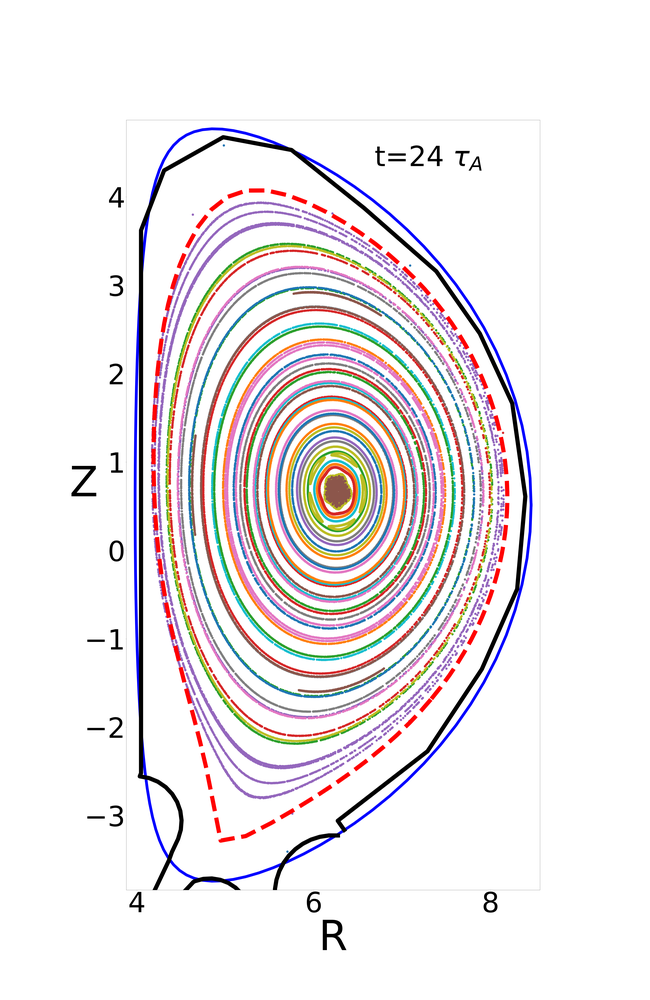}
    \caption{Poincare cross section plot of the magnetic field at the
      end of the $3000\times$ simulation. A lack of significant activity is seen in the
      field during the rapid radiative TQ. The Alfv\'en time is
      $\tau_A = 10.1 \mu$s. }
    \label{fig3000xpoincare}
\end{figure}

This section compares the $300\times$ results against that of a higher density, $3000\times$ the reference density.
The thermal and current loss rates are shown in Figure \ref{fig10}.
The analysis in Section \ref{sec:physics-regimes} predicted that, due to the
quadratic scaling in the internal power balance with the injection
density, a large enough injection ratio would lead to a rapid TQ
before major MHD modes destroy the flux surfaces.
The lack of MHD activity is seen in Figure \ref{fig3000xpoincare}.
With a density
$10\times$ higher than that of the $300\times$ case, it is not
surprising then that the TQ here occurs on a timescale 100 times
shorter.  Since this TQ is due entirely to radiation losses and not
boundary losses, and because no noteworthy MHD activity develops before the TQ is over, the CQ
therefore proceeds relatively slowly.
As the
temperature drops, the increased resistivity does lead to a much faster CQ
than in an unmitigated scenario.

\begin{figure}
    \centering
    \includegraphics[scale=0.35]{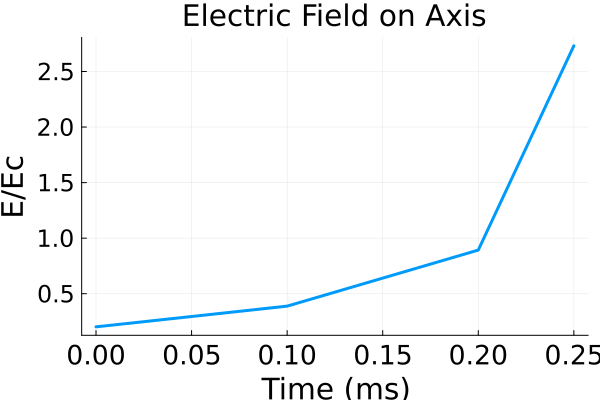}
    \caption{The parallel electric field at r=0 compared to the
      Connor-Hastie critical field value over time, for the
      $3000\times$ simulation. The critical field $E_c$ is exceeded
      during the rapid radiative TQ, suggesting runaway electrons are
      generated in this case.}
    \label{figEfield3000}
\end{figure}

The ratio of the parallel electric field to the Connor-Hastie critical field is shown in Figure \ref{figEfield3000} for the $3000\times$ case.
Unlike the $300\times$ case, the critical field is exceeded on axis,
suggesting the possibility of undesired generation of runaways.

This reference case confirms the anticipation that, as the injection density
gets too high, the TQ and CQ start to misalign again, and both can
become much shorter.  Although TQ poses less a challenge because
Bremsstrahlung radiation spreads the power load on the entire first
wall, a too short CQ could create unacceptable high electromagnetic
force loading in the blanket modules and vacuum vessel.

\section{Conclusion}\label{sec5}
We have demonstrated via MHD simulations with the PIXIE3D code that a
TQ can be modelled by including not only plasma conduction to the
chamber wall, but also a sheath outflow as well as Bremsstrahlung
losses.  The purpose of this study is a proof-of-concept that if
collisional conditions can be achieved with a high-density hydrogen
injection for disruption mitigation, the thermal quench and current quench
can be aligned on comparable time scales. This can simultaneously
overcome the plasma thermal load challenge in a fast TQ and avoid
runaway electron acceleration.  Another added benefit is that via
dilutional cooling at the very onset of disruption, the impacting
energy of the ions can be controlled to less than 100~eV, so wall
impurity production by physical sputtering is inhibited on tungsten
divertors and first wall, removing an important impurity source that
could radiatively clamp the plasma temperature to a low value for
which runaway electrons become a severe concern.  Specifically for the
ITER case we have simulated, while the unmitigated timescale of the TQ
is estimated to be around 1 ms or less, this can be prolonged by more
than an order of magnitude using a low-Z hydrogen injection.  An
injection density of $300\times$ the reference density is somewhat
arbitrary, but it does indicate that there is an optimal density
regime where parallel thermal losses are halted but radiation losses
are not too extreme.  The existence of such an optimal density regime
is further bolstered by the findings that if the density is too large,
radiation losses lead to a rapid TQ, whereas if the density is not
sufficiently high, parallel thermal conductivity also leads to a fast
TQ.  In the optimal density regime, the plasma is collisional and the
TQ will occur over a similar timescale as the CQ, meaning that the
plasma will stay closer to the Ohmic current regime and the
generation of runaway electrons is inhibited.

\begin{acknowledgments}
  This work was supported by the U.S. Department of Energy Office of
  Fusion Energy Sciences and Office of Advanced Scientific Computing
  Research through the Tokamak Disruption Simulation (TDS) SciDAC
  project, and the Base Fusion Theory Program at Los Alamos National
  Laboratory (LANL) under contract No. 89233218CNA000001.  This
  research used resources of the National Energy Research Scientific
  Computing Center, a DOE Office of Science User Facility supported by
  the Office of Science of the U.S. Department of Energy under
  Contract No. DE-AC02-05CH11231 using NERSC award FES-ERCAP0032298
  and LANL Institutional Computing Program, which is supported by the
  U.S. Department of Energy National Nuclear Security Administration
  under Contract No. 89233218CNA000001.
\end{acknowledgments}

\section*{Data Availability}

The data that support the findings of this study are available upon reasonable request.

\appendix

\section{Logical-to-physical coordinate map}\label{app:iter-mesh}

The ITER simulations conducted in this work use a logical coordinate system $\left(r,\theta,\phi\right)\,\in\,\left[0,1\right]\times\left[0,2\pi\right)\times\left[0,2\pi\right)$.
The logical coordinates are mapped to cylindrical coordinates using the transformation:
\begin{align*}
R\left(r,\theta,\phi\right)=& R_m + (R_0 -R_m) r + a r \cos{\left[\theta+\arcsin{\left(\delta \, r^2 \sin{\theta}\right)}\right]} ,\\
Z\left(r,\theta,\phi\right)=& Z_m + (Z_0 - Z_m) r + a \left(r\kappa + (1-r)\kappa_s\right) r \sin{\left[\theta+\zeta \,r^2 \sin\left(2\theta\right)\right]},\\
\phi_c\left(\phi\right)=& -\phi.
\end{align*}
The shaping parameters for the ITER experiment are: minor radius $a=2.24$m, geometric axis $R_0=6.219577546$m, $Z_0=0.5143555944$m, magnetic axis $R_m=6.341952203$m, $Z_m=0.6327986088$m, triangularity $\delta=0.6$, elongation $\kappa=1.9$, elongation at the magnetic axis $\kappa_s=1.35$, and squareness $\zeta=0.06$.
We emphasize that the cylindrical angle $\phi_c$ rotates about the origin in the opposite direction as the toroidal angle $\phi$.

\section{Simulation without sheath boundary condition and Bremsstrahlung losses produces unphysically long quenches \label{sec44}}
Instead of the sheath boundary used in the three prototypical
simulation studies, this simulation will use a conventional no-flow
boundary at $r=1$, with the same collisional $300\times$ transport
coefficients as Section \ref{sec41}.  In addition, no Bremsstrahlung
is included ($P_{B0}=0$), but there is Joule and viscous heating.
Therefore, the only energy loss mechanism is through heat conduction
at the wall.  The simulation ended after 20,000 Alfv\'en times due to
computational expense, and no other interesting physical phenomenon
was expected apart from the slow dissipation of the core plasma.

\begin{figure}
    \centering
    \subfloat[Equilibrium topology, after 0.16 ms.]{%
    \includegraphics[scale=0.2]{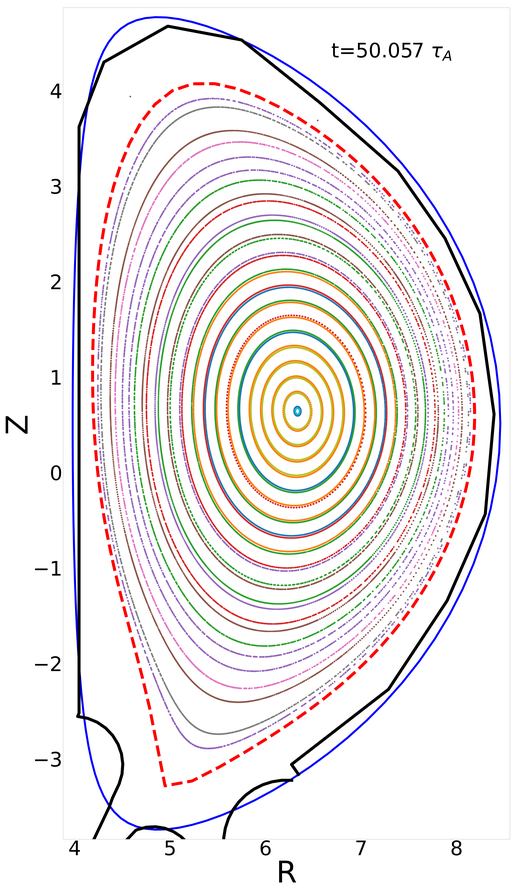}\label{fig2a}}\quad
    \subfloat[1-1 kink mode, after 14.08 ms.]{%
    \includegraphics[scale=0.2]{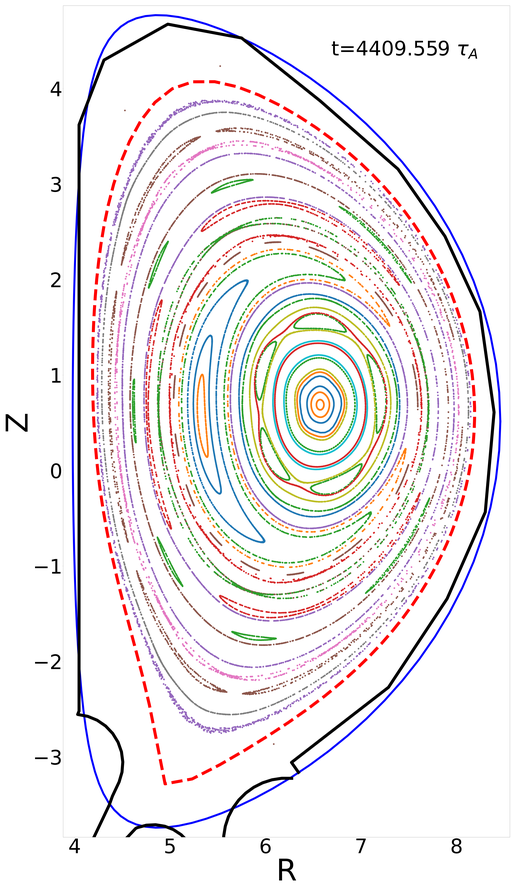}\label{fig2b}}\\
    \subfloat[Breakup of kink mode,  after 15.85 ms.]{%
    \includegraphics[scale=0.2]{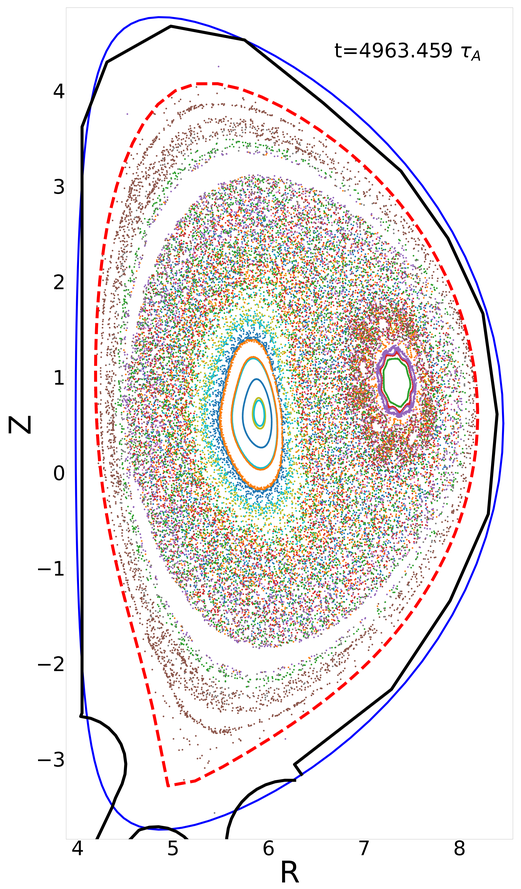}\label{fig2c}}\quad
    \subfloat[Core dissipation, after 62.19 ms.]{%
    \includegraphics[scale=0.2]{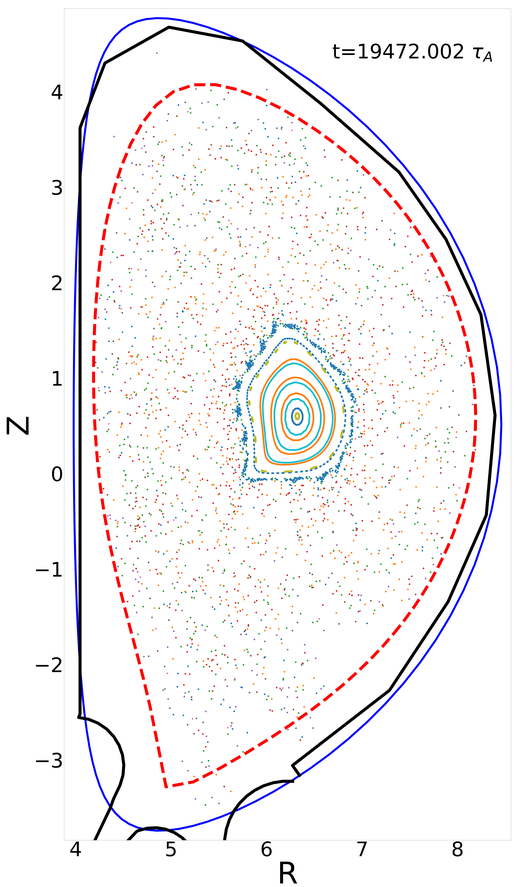}\label{fig2d}}
    \caption{Magnetic topology at various times throughout the sheathless simulation. The Alfv\'en time is $\tau_A = 3.2 \mu$s.}
    \label{fig2}
\end{figure}

\begin{figure}
    \centering
    \includegraphics[scale=0.25]{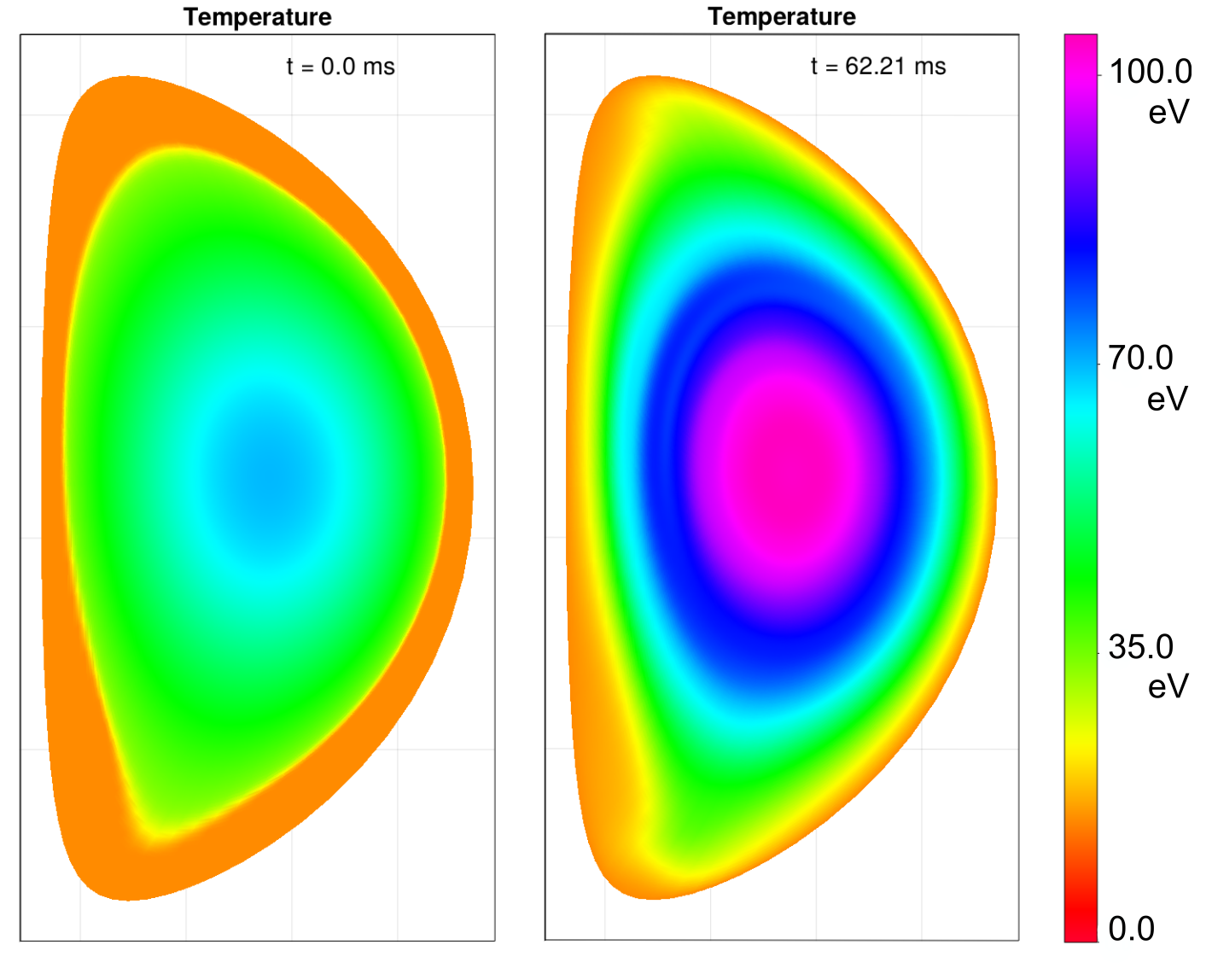}
    \caption{Initial (t = 0 ms) and final (t = 62 ms) temperature profiles for the sheathless simulation, toroidally averaged. Due to Joule heating and weak conduction into the walls, the plasma has net positive heating. For this simulation only, the temperature has a non-homogeneous Dirichlet condition at the wall fixed to 13 eV.}
    \label{fig3}
\end{figure}

The magnetic field topology is shown in Figure \ref{fig2} and the initial and final temperature profiles are shown in Figure \ref{fig3}.
Without Bremsstrahlung depleting energy in the core, the 1-1 kink mode takes significantly longer to develop compared to that of Figure \ref{fig6}.
Once disruption occurs, there is a similar topological evolution of closed flux surfaces into open field lines.
However, instead of rapidly dissipating and then recovering, the closed surfaces dissipate slowly suggesting a long CQ.

\begin{figure}
    \centering
    \includegraphics[scale=0.5]{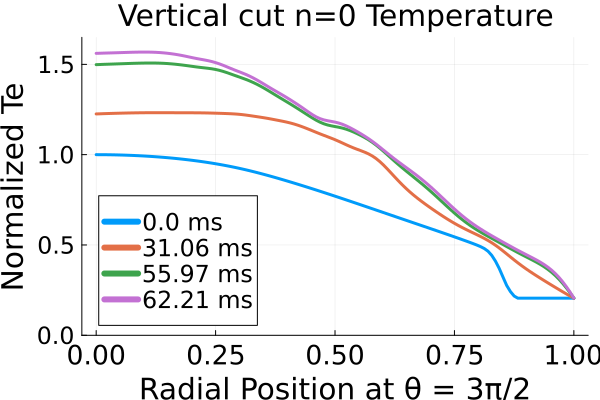}
    \caption{Radial profile for the temperature from the magnetic axis (r=0) to the divertor (r=1) for the sheathless simulation. Conduction into the wall is limited by the shallow gradient.}
    \label{fig4}
\end{figure}

\begin{figure}
    \centering
    \includegraphics[scale=0.5]{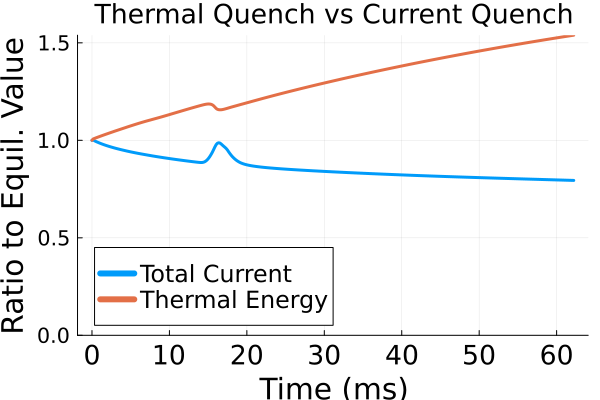}
    \caption{Time scales of the thermal and current quenches are shown via the evolution of the total thermal energy and total current respectively, for the sheathless $300\times$ simulation. The current spike occurs approximately when the core reforms following the breakup of the kink mode, see Figure \ref{fig2c}.}
    \label{fig5}
\end{figure}

The radial temperature profile shown in Figure \ref{fig4} has a modest gradient at the wall, which dictates the conduction losses.
This results in insufficient energy losses and no TQ is observed, in contrast to the simulation with a sheath (Section \ref{sec41}).
Figure \ref{fig5} shows the total thermal energy and total current integrated across the entire domain, where the CQ can be observed to occur over a sufficiently long timescale (with a current spike around 16 ms corresponding to when the core re-heals following the breakup of the kink mode), but heating sources dominate over the modest wall losses and thermal energy grows over time.
The lack of a thermal quench highlights the need for more physical energy loss mechanisms for the MHD simulations to be able to reproduce experimental signatures.
The conclusion is that conduction losses alone are inadequate for the modelling of a TQ.

%\nocite{*}

\bibliographystyle{apsrev4-2}
\bibliography{TQmitigation}

\providecommand{\noopsort}[1]{}\providecommand{\singleletter}[1]{#1}%
\begin{thebibliography}{53}%
\makeatletter
\providecommand \@ifxundefined [1]{%
 \@ifx{#1\undefined}
}%
\providecommand \@ifnum [1]{%
 \ifnum #1\expandafter \@firstoftwo
 \else \expandafter \@secondoftwo
 \fi
}%
\providecommand \@ifx [1]{%
 \ifx #1\expandafter \@firstoftwo
 \else \expandafter \@secondoftwo
 \fi
}%
\providecommand \natexlab [1]{#1}%
\providecommand \enquote  [1]{``#1''}%
\providecommand \bibnamefont  [1]{#1}%
\providecommand \bibfnamefont [1]{#1}%
\providecommand \citenamefont [1]{#1}%
\providecommand \href@noop [0]{\@secondoftwo}%
\providecommand \href [0]{\begingroup \@sanitize@url \@href}%
\providecommand \@href[1]{\@@startlink{#1}\@@href}%
\providecommand \@@href[1]{\endgroup#1\@@endlink}%
\providecommand \@sanitize@url [0]{\catcode `\\12\catcode `\$12\catcode
  `\&12\catcode `\#12\catcode `\^12\catcode `\_12\catcode `\%12\relax}%
\providecommand \@@startlink[1]{}%
\providecommand \@@endlink[0]{}%
\providecommand \url  [0]{\begingroup\@sanitize@url \@url }%
\providecommand \@url [1]{\endgroup\@href {#1}{\urlprefix }}%
\providecommand \urlprefix  [0]{URL }%
\providecommand \Eprint [0]{\href }%
\providecommand \doibase [0]{https://doi.org/}%
\providecommand \selectlanguage [0]{\@gobble}%
\providecommand \bibinfo  [0]{\@secondoftwo}%
\providecommand \bibfield  [0]{\@secondoftwo}%
\providecommand \translation [1]{[#1]}%
\providecommand \BibitemOpen [0]{}%
\providecommand \bibitemStop [0]{}%
\providecommand \bibitemNoStop [0]{.\EOS\space}%
\providecommand \EOS [0]{\spacefactor3000\relax}%
\providecommand \BibitemShut  [1]{\csname bibitem#1\endcsname}%
\let\auto@bib@innerbib\@empty
%</preamble>
\bibitem [{\citenamefont {Hender}\ \emph {et~al.}(2007)\citenamefont {Hender},
  \citenamefont {Wesley}, \citenamefont {Bialek}, \citenamefont {Bondeson},
  \citenamefont {Boozer}, \citenamefont {Buttery}, \citenamefont {Garofalo},
  \citenamefont {Goodman}, \citenamefont {Granetz}, \citenamefont {Gribov}
  \emph {et~al.}}]{Hender:2007}%
  \BibitemOpen
  \bibfield  {author} {\bibinfo {author} {\bibfnamefont {T.}~\bibnamefont
  {Hender}}, \bibinfo {author} {\bibfnamefont {J.}~\bibnamefont {Wesley}},
  \bibinfo {author} {\bibfnamefont {J.}~\bibnamefont {Bialek}}, \bibinfo
  {author} {\bibfnamefont {A.}~\bibnamefont {Bondeson}}, \bibinfo {author}
  {\bibfnamefont {A.}~\bibnamefont {Boozer}}, \bibinfo {author} {\bibfnamefont
  {R.}~\bibnamefont {Buttery}}, \bibinfo {author} {\bibfnamefont
  {A.}~\bibnamefont {Garofalo}}, \bibinfo {author} {\bibfnamefont
  {T.}~\bibnamefont {Goodman}}, \bibinfo {author} {\bibfnamefont
  {R.}~\bibnamefont {Granetz}}, \bibinfo {author} {\bibfnamefont
  {Y.}~\bibnamefont {Gribov}}, \emph {et~al.},\ }\href@noop {} {\bibfield
  {journal} {\bibinfo  {journal} {Nuclear fusion}\ }\textbf {\bibinfo {volume}
  {47}},\ \bibinfo {pages} {S128} (\bibinfo {year} {2007})}\BibitemShut
  {NoStop}%
\bibitem [{\citenamefont {Nedospasov}(2008)}]{nedospasov2008thermal}%
  \BibitemOpen
  \bibfield  {author} {\bibinfo {author} {\bibfnamefont {A.}~\bibnamefont
  {Nedospasov}},\ }\href@noop {} {\bibfield  {journal} {\bibinfo  {journal}
  {Nuclear fusion}\ }\textbf {\bibinfo {volume} {48}},\ \bibinfo {pages}
  {032002} (\bibinfo {year} {2008})}\BibitemShut {NoStop}%
\bibitem [{\citenamefont {Lehnen}\ \emph {et~al.}(2015)\citenamefont {Lehnen},
  \citenamefont {Aleynikova}, \citenamefont {Aleynikov}, \citenamefont
  {Campbell}, \citenamefont {Drewelow}, \citenamefont {Eidietis}, \citenamefont
  {Gasparyan}, \citenamefont {Granetz}, \citenamefont {Gribov}, \citenamefont
  {Hartmann} \emph {et~al.}}]{lehnen2015disruptions}%
  \BibitemOpen
  \bibfield  {author} {\bibinfo {author} {\bibfnamefont {M.}~\bibnamefont
  {Lehnen}}, \bibinfo {author} {\bibfnamefont {K.}~\bibnamefont {Aleynikova}},
  \bibinfo {author} {\bibfnamefont {P.}~\bibnamefont {Aleynikov}}, \bibinfo
  {author} {\bibfnamefont {D.}~\bibnamefont {Campbell}}, \bibinfo {author}
  {\bibfnamefont {P.}~\bibnamefont {Drewelow}}, \bibinfo {author}
  {\bibfnamefont {N.}~\bibnamefont {Eidietis}}, \bibinfo {author}
  {\bibfnamefont {Y.}~\bibnamefont {Gasparyan}}, \bibinfo {author}
  {\bibfnamefont {R.}~\bibnamefont {Granetz}}, \bibinfo {author} {\bibfnamefont
  {Y.}~\bibnamefont {Gribov}}, \bibinfo {author} {\bibfnamefont
  {N.}~\bibnamefont {Hartmann}}, \emph {et~al.},\ }\href@noop {} {\bibfield
  {journal} {\bibinfo  {journal} {Journal of Nuclear materials}\ }\textbf
  {\bibinfo {volume} {463}},\ \bibinfo {pages} {39} (\bibinfo {year}
  {2015})}\BibitemShut {NoStop}%
\bibitem [{\citenamefont {Shiraki}\ \emph {et~al.}(2016)\citenamefont
  {Shiraki}, \citenamefont {Commaux}, \citenamefont {Baylor}, \citenamefont
  {Eidietis}, \citenamefont {Hollmann}, \citenamefont {Lasnier},\ and\
  \citenamefont {Moyer}}]{shiraki2016thermal}%
  \BibitemOpen
  \bibfield  {author} {\bibinfo {author} {\bibfnamefont {D.}~\bibnamefont
  {Shiraki}}, \bibinfo {author} {\bibfnamefont {N.}~\bibnamefont {Commaux}},
  \bibinfo {author} {\bibfnamefont {L.}~\bibnamefont {Baylor}}, \bibinfo
  {author} {\bibfnamefont {N.}~\bibnamefont {Eidietis}}, \bibinfo {author}
  {\bibfnamefont {E.}~\bibnamefont {Hollmann}}, \bibinfo {author}
  {\bibfnamefont {C.}~\bibnamefont {Lasnier}},\ and\ \bibinfo {author}
  {\bibfnamefont {R.}~\bibnamefont {Moyer}},\ }\href@noop {} {\bibfield
  {journal} {\bibinfo  {journal} {Physics of Plasmas}\ }\textbf {\bibinfo
  {volume} {23}} (\bibinfo {year} {2016})}\BibitemShut {NoStop}%
\bibitem [{\citenamefont {Commaux}\ \emph {et~al.}(2016)\citenamefont
  {Commaux}, \citenamefont {Shiraki}, \citenamefont {Baylor}, \citenamefont
  {Hollmann}, \citenamefont {Eidietis}, \citenamefont {Lasnier}, \citenamefont
  {Moyer}, \citenamefont {Jernigan}, \citenamefont {Meitner}, \citenamefont
  {Combs} \emph {et~al.}}]{commaux2016first}%
  \BibitemOpen
  \bibfield  {author} {\bibinfo {author} {\bibfnamefont {N.}~\bibnamefont
  {Commaux}}, \bibinfo {author} {\bibfnamefont {D.}~\bibnamefont {Shiraki}},
  \bibinfo {author} {\bibfnamefont {L.~R.}\ \bibnamefont {Baylor}}, \bibinfo
  {author} {\bibfnamefont {E.}~\bibnamefont {Hollmann}}, \bibinfo {author}
  {\bibfnamefont {N.}~\bibnamefont {Eidietis}}, \bibinfo {author}
  {\bibfnamefont {C.}~\bibnamefont {Lasnier}}, \bibinfo {author} {\bibfnamefont
  {R.}~\bibnamefont {Moyer}}, \bibinfo {author} {\bibfnamefont
  {T.}~\bibnamefont {Jernigan}}, \bibinfo {author} {\bibfnamefont
  {S.}~\bibnamefont {Meitner}}, \bibinfo {author} {\bibfnamefont {S.~K.}\
  \bibnamefont {Combs}}, \emph {et~al.},\ }\href@noop {} {\bibfield  {journal}
  {\bibinfo  {journal} {Nuclear Fusion}\ }\textbf {\bibinfo {volume} {56}},\
  \bibinfo {pages} {046007} (\bibinfo {year} {2016})}\BibitemShut {NoStop}%
\bibitem [{\citenamefont {McDevitt}\ \emph {et~al.}(2022)\citenamefont
  {McDevitt}, \citenamefont {Tang}, \citenamefont {Fontes}, \citenamefont
  {Sharma},\ and\ \citenamefont {Chung}}]{mcdevitt2022constraint}%
  \BibitemOpen
  \bibfield  {author} {\bibinfo {author} {\bibfnamefont {C.~J.}\ \bibnamefont
  {McDevitt}}, \bibinfo {author} {\bibfnamefont {X.}~\bibnamefont {Tang}},
  \bibinfo {author} {\bibfnamefont {C.}~\bibnamefont {Fontes}}, \bibinfo
  {author} {\bibfnamefont {P.}~\bibnamefont {Sharma}},\ and\ \bibinfo {author}
  {\bibfnamefont {H.-K.}\ \bibnamefont {Chung}},\ }\href@noop {} {\bibfield
  {journal} {\bibinfo  {journal} {Nuclear Fusion}\ }\textbf {\bibinfo {volume}
  {62}},\ \bibinfo {pages} {112004} (\bibinfo {year} {2022})}\BibitemShut
  {NoStop}%
\bibitem [{\citenamefont {Dreicer}(1959)}]{Dreicer59}%
  \BibitemOpen
  \bibfield  {author} {\bibinfo {author} {\bibfnamefont {H.}~\bibnamefont
  {Dreicer}},\ }\href {https://doi.org/10.1103/PhysRev.115.238} {\bibfield
  {journal} {\bibinfo  {journal} {Phys. Rev.}\ }\textbf {\bibinfo {volume}
  {115}},\ \bibinfo {pages} {238} (\bibinfo {year} {1959})}\BibitemShut
  {NoStop}%
\bibitem [{\citenamefont {Connor}\ and\ \citenamefont
  {Hastie}(1975)}]{connor-hastie-NF75}%
  \BibitemOpen
  \bibfield  {author} {\bibinfo {author} {\bibfnamefont {J.}~\bibnamefont
  {Connor}}\ and\ \bibinfo {author} {\bibfnamefont {R.}~\bibnamefont
  {Hastie}},\ }\href {http://stacks.iop.org/0029-5515/15/i=3/a=007} {\bibfield
  {journal} {\bibinfo  {journal} {Nuclear Fusion}\ }\textbf {\bibinfo {volume}
  {15}},\ \bibinfo {pages} {415} (\bibinfo {year} {1975})}\BibitemShut
  {NoStop}%
\bibitem [{\citenamefont {Guo}\ \emph {et~al.}(2017)\citenamefont {Guo},
  \citenamefont {McDevitt},\ and\ \citenamefont {Tang}}]{guo2017phase}%
  \BibitemOpen
  \bibfield  {author} {\bibinfo {author} {\bibfnamefont {Z.}~\bibnamefont
  {Guo}}, \bibinfo {author} {\bibfnamefont {C.~J.}\ \bibnamefont {McDevitt}},\
  and\ \bibinfo {author} {\bibfnamefont {X.-Z.}\ \bibnamefont {Tang}},\
  }\href@noop {} {\bibfield  {journal} {\bibinfo  {journal} {Plasma Physics and
  Controlled Fusion}\ }\textbf {\bibinfo {volume} {59}},\ \bibinfo {pages}
  {044003} (\bibinfo {year} {2017})}\BibitemShut {NoStop}%
\bibitem [{\citenamefont {Jayakumar}\ \emph {et~al.}(1993)\citenamefont
  {Jayakumar}, \citenamefont {Fleischmann},\ and\ \citenamefont
  {Zweben}}]{Jayakumar:1993}%
  \BibitemOpen
  \bibfield  {author} {\bibinfo {author} {\bibfnamefont {R.}~\bibnamefont
  {Jayakumar}}, \bibinfo {author} {\bibfnamefont {H.}~\bibnamefont
  {Fleischmann}},\ and\ \bibinfo {author} {\bibfnamefont {S.}~\bibnamefont
  {Zweben}},\ }\href@noop {} {\bibfield  {journal} {\bibinfo  {journal}
  {Physics Letters A}\ }\textbf {\bibinfo {volume} {172}},\ \bibinfo {pages}
  {447} (\bibinfo {year} {1993})}\BibitemShut {NoStop}%
\bibitem [{\citenamefont {Rosenbluth}\ and\ \citenamefont
  {Putvinski}(1997)}]{Rosenbluth97}%
  \BibitemOpen
  \bibfield  {author} {\bibinfo {author} {\bibfnamefont {M.}~\bibnamefont
  {Rosenbluth}}\ and\ \bibinfo {author} {\bibfnamefont {S.}~\bibnamefont
  {Putvinski}},\ }\href {http://stacks.iop.org/0029-5515/37/i=10/a=I03}
  {\bibfield  {journal} {\bibinfo  {journal} {Nuclear Fusion}\ }\textbf
  {\bibinfo {volume} {37}},\ \bibinfo {pages} {1355} (\bibinfo {year}
  {1997})}\BibitemShut {NoStop}%
\bibitem [{\citenamefont {Hesslow}\ \emph {et~al.}(2017)\citenamefont
  {Hesslow}, \citenamefont {Embr\'eus}, \citenamefont {Stahl}, \citenamefont
  {DuBois}, \citenamefont {Papp}, \citenamefont {Newton},\ and\ \citenamefont
  {F\"ul\"op}}]{hesslow-etal-prl-2017}%
  \BibitemOpen
  \bibfield  {author} {\bibinfo {author} {\bibfnamefont {L.}~\bibnamefont
  {Hesslow}}, \bibinfo {author} {\bibfnamefont {O.}~\bibnamefont {Embr\'eus}},
  \bibinfo {author} {\bibfnamefont {A.}~\bibnamefont {Stahl}}, \bibinfo
  {author} {\bibfnamefont {T.~C.}\ \bibnamefont {DuBois}}, \bibinfo {author}
  {\bibfnamefont {G.}~\bibnamefont {Papp}}, \bibinfo {author} {\bibfnamefont
  {S.~L.}\ \bibnamefont {Newton}},\ and\ \bibinfo {author} {\bibfnamefont
  {T.}~\bibnamefont {F\"ul\"op}},\ }\href
  {https://doi.org/10.1103/PhysRevLett.118.255001} {\bibfield  {journal}
  {\bibinfo  {journal} {Phys. Rev. Lett.}\ }\textbf {\bibinfo {volume} {118}},\
  \bibinfo {pages} {255001} (\bibinfo {year} {2017})}\BibitemShut {NoStop}%
\bibitem [{\citenamefont {McDevitt}\ \emph {et~al.}(2019)\citenamefont
  {McDevitt}, \citenamefont {Guo},\ and\ \citenamefont {Tang}}]{McDevitt-2019}%
  \BibitemOpen
  \bibfield  {author} {\bibinfo {author} {\bibfnamefont {C.~J.}\ \bibnamefont
  {McDevitt}}, \bibinfo {author} {\bibfnamefont {Z.}~\bibnamefont {Guo}},\ and\
  \bibinfo {author} {\bibfnamefont {X.-Z.}\ \bibnamefont {Tang}},\ }\href
  {https://doi.org/10.1088/1361-6587/ab0d6d} {\bibfield  {journal} {\bibinfo
  {journal} {Plasma Physics and Controlled Fusion}\ }\textbf {\bibinfo {volume}
  {61}},\ \bibinfo {pages} {054008} (\bibinfo {year} {2019})}\BibitemShut
  {NoStop}%
\bibitem [{\citenamefont {Artola}\ \emph {et~al.}(2022)\citenamefont {Artola},
  \citenamefont {Loarte}, \citenamefont {Hoelzl}, \citenamefont {Lehnen},
  \citenamefont {Schwarz}, \citenamefont {Team} \emph
  {et~al.}}]{artola2022non}%
  \BibitemOpen
  \bibfield  {author} {\bibinfo {author} {\bibfnamefont {F.}~\bibnamefont
  {Artola}}, \bibinfo {author} {\bibfnamefont {A.}~\bibnamefont {Loarte}},
  \bibinfo {author} {\bibfnamefont {M.}~\bibnamefont {Hoelzl}}, \bibinfo
  {author} {\bibfnamefont {M.}~\bibnamefont {Lehnen}}, \bibinfo {author}
  {\bibfnamefont {N.}~\bibnamefont {Schwarz}}, \bibinfo {author} {\bibfnamefont
  {J.}~\bibnamefont {Team}}, \emph {et~al.},\ }\href@noop {} {\bibfield
  {journal} {\bibinfo  {journal} {Nuclear Fusion}\ }\textbf {\bibinfo {volume}
  {62}},\ \bibinfo {pages} {056023} (\bibinfo {year} {2022})}\BibitemShut
  {NoStop}%
\bibitem [{\citenamefont {Paz-Soldan}\ \emph {et~al.}(2021)\citenamefont
  {Paz-Soldan}, \citenamefont {Reux}, \citenamefont {Aleynikova}, \citenamefont
  {Aleynikov}, \citenamefont {Bandaru}, \citenamefont {Beidler}, \citenamefont
  {Eidietis}, \citenamefont {Liu}, \citenamefont {Liu}, \citenamefont
  {Lvovskiy} \emph {et~al.}}]{paz2021novel}%
  \BibitemOpen
  \bibfield  {author} {\bibinfo {author} {\bibfnamefont {C.}~\bibnamefont
  {Paz-Soldan}}, \bibinfo {author} {\bibfnamefont {C.}~\bibnamefont {Reux}},
  \bibinfo {author} {\bibfnamefont {K.}~\bibnamefont {Aleynikova}}, \bibinfo
  {author} {\bibfnamefont {P.}~\bibnamefont {Aleynikov}}, \bibinfo {author}
  {\bibfnamefont {V.}~\bibnamefont {Bandaru}}, \bibinfo {author} {\bibfnamefont
  {M.}~\bibnamefont {Beidler}}, \bibinfo {author} {\bibfnamefont
  {N.}~\bibnamefont {Eidietis}}, \bibinfo {author} {\bibfnamefont
  {Y.}~\bibnamefont {Liu}}, \bibinfo {author} {\bibfnamefont {C.}~\bibnamefont
  {Liu}}, \bibinfo {author} {\bibfnamefont {A.}~\bibnamefont {Lvovskiy}}, \emph
  {et~al.},\ }\href@noop {} {\bibfield  {journal} {\bibinfo  {journal} {Nuclear
  Fusion}\ }\textbf {\bibinfo {volume} {61}},\ \bibinfo {pages} {116058}
  (\bibinfo {year} {2021})}\BibitemShut {NoStop}%
\bibitem [{\citenamefont {Reux}\ \emph {et~al.}(2021)\citenamefont {Reux},
  \citenamefont {Paz-Soldan}, \citenamefont {Aleynikov}, \citenamefont
  {Bandaru}, \citenamefont {Ficker}, \citenamefont {Silburn}, \citenamefont
  {Hoelzl}, \citenamefont {Jachmich}, \citenamefont {Eidietis}, \citenamefont
  {Lehnen} \emph {et~al.}}]{reux2021demonstration}%
  \BibitemOpen
  \bibfield  {author} {\bibinfo {author} {\bibfnamefont {C.}~\bibnamefont
  {Reux}}, \bibinfo {author} {\bibfnamefont {C.}~\bibnamefont {Paz-Soldan}},
  \bibinfo {author} {\bibfnamefont {P.}~\bibnamefont {Aleynikov}}, \bibinfo
  {author} {\bibfnamefont {V.}~\bibnamefont {Bandaru}}, \bibinfo {author}
  {\bibfnamefont {O.}~\bibnamefont {Ficker}}, \bibinfo {author} {\bibfnamefont
  {S.}~\bibnamefont {Silburn}}, \bibinfo {author} {\bibfnamefont
  {M.}~\bibnamefont {Hoelzl}}, \bibinfo {author} {\bibfnamefont
  {S.}~\bibnamefont {Jachmich}}, \bibinfo {author} {\bibfnamefont
  {N.}~\bibnamefont {Eidietis}}, \bibinfo {author} {\bibfnamefont
  {M.}~\bibnamefont {Lehnen}}, \emph {et~al.},\ }\href@noop {} {\bibfield
  {journal} {\bibinfo  {journal} {Physical Review Letters}\ }\textbf {\bibinfo
  {volume} {126}},\ \bibinfo {pages} {175001} (\bibinfo {year}
  {2021})}\BibitemShut {NoStop}%
\bibitem [{\citenamefont {McDevitt}\ and\ \citenamefont
  {Tang}(2023)}]{McDevitt-Tang-PRE-2023}%
  \BibitemOpen
  \bibfield  {author} {\bibinfo {author} {\bibfnamefont {C.~J.}\ \bibnamefont
  {McDevitt}}\ and\ \bibinfo {author} {\bibfnamefont {X.-Z.}\ \bibnamefont
  {Tang}},\ }\href {https://doi.org/10.1103/PhysRevE.108.L043201} {\bibfield
  {journal} {\bibinfo  {journal} {Phys. Rev. E}\ }\textbf {\bibinfo {volume}
  {108}},\ \bibinfo {pages} {L043201} (\bibinfo {year} {2023})}\BibitemShut
  {NoStop}%
\bibitem [{\citenamefont {Lively}\ \emph {et~al.}(2024)\citenamefont {Lively},
  \citenamefont {Perez}, \citenamefont {Uberuaga}, \citenamefont {Zhang},\ and\
  \citenamefont {Tang}}]{Lively-etal-NF-2024}%
  \BibitemOpen
  \bibfield  {author} {\bibinfo {author} {\bibfnamefont {M.~A.}\ \bibnamefont
  {Lively}}, \bibinfo {author} {\bibfnamefont {D.}~\bibnamefont {Perez}},
  \bibinfo {author} {\bibfnamefont {B.~P.}\ \bibnamefont {Uberuaga}}, \bibinfo
  {author} {\bibfnamefont {Y.}~\bibnamefont {Zhang}},\ and\ \bibinfo {author}
  {\bibfnamefont {X.-Z.}\ \bibnamefont {Tang}},\ }\href
  {https://doi.org/10.1088/1741-4326/ad35d5} {\bibfield  {journal} {\bibinfo
  {journal} {Nuclear Fusion}\ }\textbf {\bibinfo {volume} {64}},\ \bibinfo
  {pages} {056019} (\bibinfo {year} {2024})}\BibitemShut {NoStop}%
\bibitem [{\citenamefont {Eidietis}(2021)}]{eidietis-fst-2021}%
  \BibitemOpen
  \bibfield  {author} {\bibinfo {author} {\bibfnamefont {N.~W.}\ \bibnamefont
  {Eidietis}},\ }\href {https://doi.org/10.1080/15361055.2021.1889919}
  {\bibfield  {journal} {\bibinfo  {journal} {Fusion Science and Technology}\
  }\textbf {\bibinfo {volume} {77}},\ \bibinfo {pages} {738} (\bibinfo {year}
  {2021})},\ \Eprint
  {https://arxiv.org/abs/https://doi.org/10.1080/15361055.2021.1889919}
  {https://doi.org/10.1080/15361055.2021.1889919} \BibitemShut {NoStop}%
\bibitem [{\citenamefont {Zhang}\ \emph
  {et~al.}(2023{\natexlab{a}})\citenamefont {Zhang}, \citenamefont {Li},\ and\
  \citenamefont {Tang}}]{zhang2023cooling}%
  \BibitemOpen
  \bibfield  {author} {\bibinfo {author} {\bibfnamefont {Y.}~\bibnamefont
  {Zhang}}, \bibinfo {author} {\bibfnamefont {J.}~\bibnamefont {Li}},\ and\
  \bibinfo {author} {\bibfnamefont {X.-Z.}\ \bibnamefont {Tang}},\ }\href@noop
  {} {\bibfield  {journal} {\bibinfo  {journal} {Europhysics Letters}\ }\textbf
  {\bibinfo {volume} {141}},\ \bibinfo {pages} {54002} (\bibinfo {year}
  {2023}{\natexlab{a}})}\BibitemShut {NoStop}%
\bibitem [{\citenamefont {Zhang}\ \emph
  {et~al.}(2023{\natexlab{b}})\citenamefont {Zhang}, \citenamefont {Li},\ and\
  \citenamefont {Tang}}]{zhang2023electron}%
  \BibitemOpen
  \bibfield  {author} {\bibinfo {author} {\bibfnamefont {Y.}~\bibnamefont
  {Zhang}}, \bibinfo {author} {\bibfnamefont {J.}~\bibnamefont {Li}},\ and\
  \bibinfo {author} {\bibfnamefont {X.-Z.}\ \bibnamefont {Tang}},\ }\href@noop
  {} {\bibfield  {journal} {\bibinfo  {journal} {Physics of Plasmas}\ }\textbf
  {\bibinfo {volume} {30}},\ \bibinfo {pages} {092301} (\bibinfo {year}
  {2023}{\natexlab{b}})}\BibitemShut {NoStop}%
\bibitem [{\citenamefont {Li}\ \emph {et~al.}(2023)\citenamefont {Li},
  \citenamefont {Zhang},\ and\ \citenamefont {Tang}}]{li2023staged}%
  \BibitemOpen
  \bibfield  {author} {\bibinfo {author} {\bibfnamefont {J.}~\bibnamefont
  {Li}}, \bibinfo {author} {\bibfnamefont {Y.}~\bibnamefont {Zhang}},\ and\
  \bibinfo {author} {\bibfnamefont {X.-Z.}\ \bibnamefont {Tang}},\ }\href@noop
  {} {\bibfield  {journal} {\bibinfo  {journal} {Nuclear Fusion}\ }\textbf
  {\bibinfo {volume} {63}},\ \bibinfo {pages} {066030} (\bibinfo {year}
  {2023})}\BibitemShut {NoStop}%
\bibitem [{\citenamefont {Zhang}\ \emph {et~al.}(2024)\citenamefont {Zhang},
  \citenamefont {Li},\ and\ \citenamefont {Tang}}]{zhang2024collisionless}%
  \BibitemOpen
  \bibfield  {author} {\bibinfo {author} {\bibfnamefont {Y.}~\bibnamefont
  {Zhang}}, \bibinfo {author} {\bibfnamefont {J.}~\bibnamefont {Li}},\ and\
  \bibinfo {author} {\bibfnamefont {X.-Z.}\ \bibnamefont {Tang}},\ }\href@noop
  {} {\bibfield  {journal} {\bibinfo  {journal} {Scientific Reports}\ }\textbf
  {\bibinfo {volume} {14}},\ \bibinfo {pages} {23448} (\bibinfo {year}
  {2024})}\BibitemShut {NoStop}%
\bibitem [{\citenamefont {Chac{\'o}n}(2004)}]{chacon2004non}%
  \BibitemOpen
  \bibfield  {author} {\bibinfo {author} {\bibfnamefont {L.}~\bibnamefont
  {Chac{\'o}n}},\ }\href@noop {} {\bibfield  {journal} {\bibinfo  {journal}
  {Computer Physics Communications}\ }\textbf {\bibinfo {volume} {163}},\
  \bibinfo {pages} {143} (\bibinfo {year} {2004})}\BibitemShut {NoStop}%
\bibitem [{\citenamefont {Chac{\'o}n}(2008)}]{chacon2008optimal}%
  \BibitemOpen
  \bibfield  {author} {\bibinfo {author} {\bibfnamefont {L.}~\bibnamefont
  {Chac{\'o}n}},\ }\href@noop {} {\bibfield  {journal} {\bibinfo  {journal}
  {Physics of Plasmas}\ }\textbf {\bibinfo {volume} {15}} (\bibinfo {year}
  {2008})}\BibitemShut {NoStop}%
\bibitem [{\citenamefont {Braginskii}(1965)}]{braginskii1965transport}%
  \BibitemOpen
  \bibfield  {author} {\bibinfo {author} {\bibfnamefont {S.}~\bibnamefont
  {Braginskii}},\ }\href@noop {} {\bibfield  {journal} {\bibinfo  {journal}
  {Reviews of plasma physics}\ }\textbf {\bibinfo {volume} {1}},\ \bibinfo
  {pages} {205} (\bibinfo {year} {1965})}\BibitemShut {NoStop}%
\bibitem [{\citenamefont {Tang}\ and\ \citenamefont
  {Guo}(2017)}]{tang-guo-nme-2017}%
  \BibitemOpen
  \bibfield  {author} {\bibinfo {author} {\bibfnamefont {X.-Z.}\ \bibnamefont
  {Tang}}\ and\ \bibinfo {author} {\bibfnamefont {Z.}~\bibnamefont {Guo}},\
  }\href {https://doi.org/http://dx.doi.org/10.1016/j.nme.2017.05.011}
  {\bibfield  {journal} {\bibinfo  {journal} {Nuclear Materials and Energy}\ }
  (\bibinfo {year} {2017})}\BibitemShut {NoStop}%
\bibitem [{\citenamefont {Tang}\ and\ \citenamefont
  {Guo}(2016{\natexlab{a}})}]{tang-guo-pop-2016}%
  \BibitemOpen
  \bibfield  {author} {\bibinfo {author} {\bibfnamefont {X.-Z.}\ \bibnamefont
  {Tang}}\ and\ \bibinfo {author} {\bibfnamefont {Z.}~\bibnamefont {Guo}},\
  }\href {https://doi.org/http://dx.doi.org/10.1063/1.4960321} {\bibfield
  {journal} {\bibinfo  {journal} {Physics of Plasmas}\ }\textbf {\bibinfo
  {volume} {23}},\ \bibinfo {eid} {083503} (\bibinfo {year}
  {2016}{\natexlab{a}})}\BibitemShut {NoStop}%
\bibitem [{\citenamefont {Tang}\ and\ \citenamefont
  {Guo}(2015)}]{tang2015sheath}%
  \BibitemOpen
  \bibfield  {author} {\bibinfo {author} {\bibfnamefont {X.-Z.}\ \bibnamefont
  {Tang}}\ and\ \bibinfo {author} {\bibfnamefont {Z.}~\bibnamefont {Guo}},\
  }\href@noop {} {\bibfield  {journal} {\bibinfo  {journal} {Physics of
  Plasmas}\ }\textbf {\bibinfo {volume} {22}} (\bibinfo {year}
  {2015})}\BibitemShut {NoStop}%
\bibitem [{\citenamefont {Artola}\ \emph {et~al.}(2021)\citenamefont {Artola},
  \citenamefont {Loarte}, \citenamefont {Matveeva}, \citenamefont {Havlicek},
  \citenamefont {Markovic}, \citenamefont {Adamek}, \citenamefont {Cavalier},
  \citenamefont {Kripner}, \citenamefont {Huijsmans}, \citenamefont {Lehnen}
  \emph {et~al.}}]{artola2021simulations}%
  \BibitemOpen
  \bibfield  {author} {\bibinfo {author} {\bibfnamefont {F.~J.}\ \bibnamefont
  {Artola}}, \bibinfo {author} {\bibfnamefont {A.}~\bibnamefont {Loarte}},
  \bibinfo {author} {\bibfnamefont {E.}~\bibnamefont {Matveeva}}, \bibinfo
  {author} {\bibfnamefont {J.}~\bibnamefont {Havlicek}}, \bibinfo {author}
  {\bibfnamefont {T.}~\bibnamefont {Markovic}}, \bibinfo {author}
  {\bibfnamefont {J.}~\bibnamefont {Adamek}}, \bibinfo {author} {\bibfnamefont
  {J.}~\bibnamefont {Cavalier}}, \bibinfo {author} {\bibfnamefont
  {L.}~\bibnamefont {Kripner}}, \bibinfo {author} {\bibfnamefont {G.~T.}\
  \bibnamefont {Huijsmans}}, \bibinfo {author} {\bibfnamefont {M.}~\bibnamefont
  {Lehnen}}, \emph {et~al.},\ }\href@noop {} {\bibfield  {journal} {\bibinfo
  {journal} {Plasma Physics and Controlled Fusion}\ }\textbf {\bibinfo {volume}
  {63}},\ \bibinfo {pages} {064004} (\bibinfo {year} {2021})}\BibitemShut
  {NoStop}%
\bibitem [{\citenamefont {Dekeyser}\ \emph {et~al.}(2021)\citenamefont
  {Dekeyser}, \citenamefont {Boerner}, \citenamefont {Voskoboynikov},
  \citenamefont {Rozhanksy}, \citenamefont {Senichenkov}, \citenamefont
  {Kaveeva}, \citenamefont {Veselova}, \citenamefont {Vekshina}, \citenamefont
  {Bonnin}, \citenamefont {Pitts} \emph {et~al.}}]{dekeyser2021plasma}%
  \BibitemOpen
  \bibfield  {author} {\bibinfo {author} {\bibfnamefont {W.}~\bibnamefont
  {Dekeyser}}, \bibinfo {author} {\bibfnamefont {P.}~\bibnamefont {Boerner}},
  \bibinfo {author} {\bibfnamefont {S.}~\bibnamefont {Voskoboynikov}}, \bibinfo
  {author} {\bibfnamefont {V.}~\bibnamefont {Rozhanksy}}, \bibinfo {author}
  {\bibfnamefont {I.}~\bibnamefont {Senichenkov}}, \bibinfo {author}
  {\bibfnamefont {L.}~\bibnamefont {Kaveeva}}, \bibinfo {author} {\bibfnamefont
  {I.}~\bibnamefont {Veselova}}, \bibinfo {author} {\bibfnamefont
  {E.}~\bibnamefont {Vekshina}}, \bibinfo {author} {\bibfnamefont
  {X.}~\bibnamefont {Bonnin}}, \bibinfo {author} {\bibfnamefont
  {R.}~\bibnamefont {Pitts}}, \emph {et~al.},\ }\href@noop {} {\bibfield
  {journal} {\bibinfo  {journal} {Nuclear Materials and Energy}\ }\textbf
  {\bibinfo {volume} {27}},\ \bibinfo {pages} {100999} (\bibinfo {year}
  {2021})}\BibitemShut {NoStop}%
\bibitem [{\citenamefont {Tang}\ and\ \citenamefont
  {Guo}(2016{\natexlab{b}})}]{tang-guo-pop-2016L}%
  \BibitemOpen
  \bibfield  {author} {\bibinfo {author} {\bibfnamefont {X.-Z.}\ \bibnamefont
  {Tang}}\ and\ \bibinfo {author} {\bibfnamefont {Z.}~\bibnamefont {Guo}},\
  }\href {https://doi.org/http://dx.doi.org/10.1063/1.4971808} {\bibfield
  {journal} {\bibinfo  {journal} {Physics of Plasmas}\ }\textbf {\bibinfo
  {volume} {23}},\ \bibinfo {eid} {120701} (\bibinfo {year}
  {2016}{\natexlab{b}})}\BibitemShut {NoStop}%
\bibitem [{\citenamefont {Li}\ \emph {et~al.}(2022{\natexlab{a}})\citenamefont
  {Li}, \citenamefont {Srinivasan}, \citenamefont {Zhang},\ and\ \citenamefont
  {Tang}}]{Li-etal-prl-2022}%
  \BibitemOpen
  \bibfield  {author} {\bibinfo {author} {\bibfnamefont {Y.}~\bibnamefont
  {Li}}, \bibinfo {author} {\bibfnamefont {B.}~\bibnamefont {Srinivasan}},
  \bibinfo {author} {\bibfnamefont {Y.}~\bibnamefont {Zhang}},\ and\ \bibinfo
  {author} {\bibfnamefont {X.-Z.}\ \bibnamefont {Tang}},\ }\href
  {https://doi.org/10.1103/PhysRevLett.128.085002} {\bibfield  {journal}
  {\bibinfo  {journal} {Phys. Rev. Lett.}\ }\textbf {\bibinfo {volume} {128}},\
  \bibinfo {pages} {085002} (\bibinfo {year} {2022}{\natexlab{a}})}\BibitemShut
  {NoStop}%
\bibitem [{\citenamefont {Li}\ \emph {et~al.}(2022{\natexlab{b}})\citenamefont
  {Li}, \citenamefont {Srinivasan}, \citenamefont {Zhang},\ and\ \citenamefont
  {Tang}}]{li2022transport}%
  \BibitemOpen
  \bibfield  {author} {\bibinfo {author} {\bibfnamefont {Y.}~\bibnamefont
  {Li}}, \bibinfo {author} {\bibfnamefont {B.}~\bibnamefont {Srinivasan}},
  \bibinfo {author} {\bibfnamefont {Y.}~\bibnamefont {Zhang}},\ and\ \bibinfo
  {author} {\bibfnamefont {X.-Z.}\ \bibnamefont {Tang}},\ }\href@noop {}
  {\bibfield  {journal} {\bibinfo  {journal} {Physics of Plasmas}\ }\textbf
  {\bibinfo {volume} {29}} (\bibinfo {year} {2022}{\natexlab{b}})}\BibitemShut
  {NoStop}%
\bibitem [{\citenamefont {Matthews}\ \emph {et~al.}(1990)\citenamefont
  {Matthews}, \citenamefont {Fielding}, \citenamefont {McCracken},
  \citenamefont {Pitcher}, \citenamefont {Stangeby},\ and\ \citenamefont
  {Ulrickson}}]{matthews1990investigation}%
  \BibitemOpen
  \bibfield  {author} {\bibinfo {author} {\bibfnamefont {G.}~\bibnamefont
  {Matthews}}, \bibinfo {author} {\bibfnamefont {S.}~\bibnamefont {Fielding}},
  \bibinfo {author} {\bibfnamefont {G.}~\bibnamefont {McCracken}}, \bibinfo
  {author} {\bibfnamefont {C.}~\bibnamefont {Pitcher}}, \bibinfo {author}
  {\bibfnamefont {P.}~\bibnamefont {Stangeby}},\ and\ \bibinfo {author}
  {\bibfnamefont {M.}~\bibnamefont {Ulrickson}},\ }\href@noop {} {\bibfield
  {journal} {\bibinfo  {journal} {Plasma Physics and Controlled Fusion}\
  }\textbf {\bibinfo {volume} {32}},\ \bibinfo {pages} {1301} (\bibinfo {year}
  {1990})}\BibitemShut {NoStop}%
\bibitem [{\citenamefont {Liu}\ \emph {et~al.}(2015)\citenamefont {Liu},
  \citenamefont {Akers}, \citenamefont {Chapman}, \citenamefont {Gribov},
  \citenamefont {Hao}, \citenamefont {Huijsmans}, \citenamefont {Kirk},
  \citenamefont {Loarte}, \citenamefont {Pinches}, \citenamefont {Reinke},
  \citenamefont {Ryan}, \citenamefont {Sun},\ and\ \citenamefont
  {Wang}}]{Liu-etal-NF-2015}%
  \BibitemOpen
  \bibfield  {author} {\bibinfo {author} {\bibfnamefont {Y.}~\bibnamefont
  {Liu}}, \bibinfo {author} {\bibfnamefont {R.}~\bibnamefont {Akers}}, \bibinfo
  {author} {\bibfnamefont {I.}~\bibnamefont {Chapman}}, \bibinfo {author}
  {\bibfnamefont {Y.}~\bibnamefont {Gribov}}, \bibinfo {author} {\bibfnamefont
  {G.}~\bibnamefont {Hao}}, \bibinfo {author} {\bibfnamefont {G.}~\bibnamefont
  {Huijsmans}}, \bibinfo {author} {\bibfnamefont {A.}~\bibnamefont {Kirk}},
  \bibinfo {author} {\bibfnamefont {A.}~\bibnamefont {Loarte}}, \bibinfo
  {author} {\bibfnamefont {S.}~\bibnamefont {Pinches}}, \bibinfo {author}
  {\bibfnamefont {M.}~\bibnamefont {Reinke}}, \bibinfo {author} {\bibfnamefont
  {D.}~\bibnamefont {Ryan}}, \bibinfo {author} {\bibfnamefont {Y.}~\bibnamefont
  {Sun}},\ and\ \bibinfo {author} {\bibfnamefont {Z.}~\bibnamefont {Wang}},\
  }\href {https://doi.org/10.1088/0029-5515/55/6/063027} {\bibfield  {journal}
  {\bibinfo  {journal} {Nuclear Fusion}\ }\textbf {\bibinfo {volume} {55}},\
  \bibinfo {pages} {063027} (\bibinfo {year} {2015})}\BibitemShut {NoStop}%
\bibitem [{\citenamefont {Liu}\ \emph {et~al.}(2021)\citenamefont {Liu},
  \citenamefont {Tang},\ and\ \citenamefont {Tang}}]{Liu-etal-SIAM-JCC-2021}%
  \BibitemOpen
  \bibfield  {author} {\bibinfo {author} {\bibfnamefont {S.}~\bibnamefont
  {Liu}}, \bibinfo {author} {\bibfnamefont {Q.}~\bibnamefont {Tang}},\ and\
  \bibinfo {author} {\bibfnamefont {X.-z.}\ \bibnamefont {Tang}},\ }\href
  {https://doi.org/10.1137/20M1385470} {\bibfield  {journal} {\bibinfo
  {journal} {SIAM Journal on Scientific Computing}\ }\textbf {\bibinfo {volume}
  {43}},\ \bibinfo {pages} {B1198} (\bibinfo {year} {2021})},\ \Eprint
  {https://arxiv.org/abs/https://doi.org/10.1137/20M1385470}
  {https://doi.org/10.1137/20M1385470} \BibitemShut {NoStop}%
\bibitem [{\citenamefont {Strauss}(2021)}]{strauss2021thermal}%
  \BibitemOpen
  \bibfield  {author} {\bibinfo {author} {\bibfnamefont {H.}~\bibnamefont
  {Strauss}},\ }\href@noop {} {\bibfield  {journal} {\bibinfo  {journal}
  {Physics of Plasmas}\ }\textbf {\bibinfo {volume} {28}} (\bibinfo {year}
  {2021})}\BibitemShut {NoStop}%
\bibitem [{\citenamefont {Artola}\ \emph {et~al.}(2024)\citenamefont {Artola},
  \citenamefont {Schwarz}, \citenamefont {Gerasimov}, \citenamefont {Loarte},\
  and\ \citenamefont {H{\"o}lzl}}]{artola2024modelling}%
  \BibitemOpen
  \bibfield  {author} {\bibinfo {author} {\bibfnamefont {F.~J.}\ \bibnamefont
  {Artola}}, \bibinfo {author} {\bibfnamefont {N.}~\bibnamefont {Schwarz}},
  \bibinfo {author} {\bibfnamefont {S.}~\bibnamefont {Gerasimov}}, \bibinfo
  {author} {\bibfnamefont {A.}~\bibnamefont {Loarte}},\ and\ \bibinfo {author}
  {\bibfnamefont {M.}~\bibnamefont {H{\"o}lzl}},\ }\href@noop {} {\bibfield
  {journal} {\bibinfo  {journal} {Plasma Physics and Controlled Fusion}\ }
  (\bibinfo {year} {2024})}\BibitemShut {NoStop}%
\bibitem [{\citenamefont {Spinicci}\ \emph {et~al.}(2023)\citenamefont
  {Spinicci}, \citenamefont {Bonfiglio}, \citenamefont {Chac{\'o}n},
  \citenamefont {Cappello},\ and\ \citenamefont
  {Veranda}}]{spinicci2023nonlinear}%
  \BibitemOpen
  \bibfield  {author} {\bibinfo {author} {\bibfnamefont {L.}~\bibnamefont
  {Spinicci}}, \bibinfo {author} {\bibfnamefont {D.}~\bibnamefont {Bonfiglio}},
  \bibinfo {author} {\bibfnamefont {L.}~\bibnamefont {Chac{\'o}n}}, \bibinfo
  {author} {\bibfnamefont {S.}~\bibnamefont {Cappello}},\ and\ \bibinfo
  {author} {\bibfnamefont {M.}~\bibnamefont {Veranda}},\ }\href@noop {}
  {\bibfield  {journal} {\bibinfo  {journal} {AIP Advances}\ }\textbf {\bibinfo
  {volume} {13}} (\bibinfo {year} {2023})}\BibitemShut {NoStop}%
\bibitem [{\citenamefont {Chac{\'o}n}\ \emph {et~al.}(tted)\citenamefont
  {Chac{\'o}n}, \citenamefont {Hamilton},\ and\ \citenamefont
  {Krasheninnikova}}]{chacon2024a}%
  \BibitemOpen
  \bibfield  {author} {\bibinfo {author} {\bibfnamefont {L.}~\bibnamefont
  {Chac{\'o}n}}, \bibinfo {author} {\bibfnamefont {J.}~\bibnamefont
  {Hamilton}},\ and\ \bibinfo {author} {\bibfnamefont {N.}~\bibnamefont
  {Krasheninnikova}},\ }\href@noop {} {\bibfield  {journal} {\bibinfo
  {journal} {Computer Physics Communications}\ } (\bibinfo {year}
  {Submitted})}\BibitemShut {NoStop}%
\bibitem [{\citenamefont {Zhdanov}(2002)}]{zhdanov2002transport}%
  \BibitemOpen
  \bibfield  {author} {\bibinfo {author} {\bibfnamefont {V.~M.}\ \bibnamefont
  {Zhdanov}},\ }\href@noop {} {\emph {\bibinfo {title} {Transport processes in
  multicomponent plasma}}}\ (\bibinfo  {publisher} {CRC Press},\ \bibinfo
  {year} {2002})\BibitemShut {NoStop}%
\bibitem [{\citenamefont {Davies}\ \emph {et~al.}(2021)\citenamefont {Davies},
  \citenamefont {Wen}, \citenamefont {Ji},\ and\ \citenamefont
  {Held}}]{davies2021transport}%
  \BibitemOpen
  \bibfield  {author} {\bibinfo {author} {\bibfnamefont {J.}~\bibnamefont
  {Davies}}, \bibinfo {author} {\bibfnamefont {H.}~\bibnamefont {Wen}},
  \bibinfo {author} {\bibfnamefont {J.-Y.}\ \bibnamefont {Ji}},\ and\ \bibinfo
  {author} {\bibfnamefont {E.~D.}\ \bibnamefont {Held}},\ }\href@noop {}
  {\bibfield  {journal} {\bibinfo  {journal} {Physics of Plasmas}\ }\textbf
  {\bibinfo {volume} {28}} (\bibinfo {year} {2021})}\BibitemShut {NoStop}%
\bibitem [{\citenamefont {Hamilton}\ and\ \citenamefont
  {Seyler}(2021)}]{hamilton2021formulation}%
  \BibitemOpen
  \bibfield  {author} {\bibinfo {author} {\bibfnamefont {J.}~\bibnamefont
  {Hamilton}}\ and\ \bibinfo {author} {\bibfnamefont {C.~E.}\ \bibnamefont
  {Seyler}},\ }\href@noop {} {\bibfield  {journal} {\bibinfo  {journal}
  {Physics of Plasmas}\ }\textbf {\bibinfo {volume} {28}} (\bibinfo {year}
  {2021})}\BibitemShut {NoStop}%
\bibitem [{\citenamefont {Hamilton}\ and\ \citenamefont
  {Seyler}(2022)}]{hamilton2022plasma}%
  \BibitemOpen
  \bibfield  {author} {\bibinfo {author} {\bibfnamefont {J.}~\bibnamefont
  {Hamilton}}\ and\ \bibinfo {author} {\bibfnamefont {C.~E.}\ \bibnamefont
  {Seyler}},\ }\href@noop {} {\bibfield  {journal} {\bibinfo  {journal}
  {Physics of Plasmas}\ }\textbf {\bibinfo {volume} {29}} (\bibinfo {year}
  {2022})}\BibitemShut {NoStop}%
\bibitem [{\citenamefont {Glasstone}\ and\ \citenamefont
  {Lovberg}(1960)}]{glasstone1960controlled}%
  \BibitemOpen
  \bibfield  {author} {\bibinfo {author} {\bibfnamefont {S.}~\bibnamefont
  {Glasstone}}\ and\ \bibinfo {author} {\bibfnamefont {R.~H.}\ \bibnamefont
  {Lovberg}},\ }\href@noop {} {\emph {\bibinfo {title} {Controlled
  thermonuclear reactions: an introduction to theory and experiment}}}\
  (\bibinfo  {publisher} {Van Nostrand},\ \bibinfo {year} {1960})\BibitemShut
  {NoStop}%
\bibitem [{\citenamefont {Delzanno}\ \emph {et~al.}(2008)\citenamefont
  {Delzanno}, \citenamefont {Chac{\'o}n},\ and\ \citenamefont
  {Finn}}]{delzanno2008electrostatic}%
  \BibitemOpen
  \bibfield  {author} {\bibinfo {author} {\bibfnamefont {G.~L.}\ \bibnamefont
  {Delzanno}}, \bibinfo {author} {\bibfnamefont {L.}~\bibnamefont
  {Chac{\'o}n}},\ and\ \bibinfo {author} {\bibfnamefont {J.~M.}\ \bibnamefont
  {Finn}},\ }\href@noop {} {\bibfield  {journal} {\bibinfo  {journal} {Physics
  of Plasmas}\ }\textbf {\bibinfo {volume} {15}} (\bibinfo {year}
  {2008})}\BibitemShut {NoStop}%
\bibitem [{\citenamefont {Stangeby}(2000)}]{stangeby-book-2000}%
  \BibitemOpen
  \bibfield  {author} {\bibinfo {author} {\bibfnamefont {P.}~\bibnamefont
  {Stangeby}},\ }\href {https://doi.org/https://doi.org/10.1201/9780367801489}
  {\emph {\bibinfo {title} {The Plasma Boundary of Magnetic Fusion Devices (1st
  ed.)}}}\ (\bibinfo  {publisher} {CRC Press},\ \bibinfo {year}
  {2000})\BibitemShut {NoStop}%
\bibitem [{\citenamefont {Taylor}(1986)}]{Taylor-prl-1986}%
  \BibitemOpen
  \bibfield  {author} {\bibinfo {author} {\bibfnamefont {J.~B.}\ \bibnamefont
  {Taylor}},\ }\href {https://doi.org/10.1103/RevModPhys.58.741} {\bibfield
  {journal} {\bibinfo  {journal} {Rev. Mod. Phys.}\ }\textbf {\bibinfo {volume}
  {58}},\ \bibinfo {pages} {741} (\bibinfo {year} {1986})}\BibitemShut
  {NoStop}%
\bibitem [{\citenamefont {Tang}\ and\ \citenamefont
  {Boozer}(2004)}]{Tang-Boozer-PoP-2004}%
  \BibitemOpen
  \bibfield  {author} {\bibinfo {author} {\bibfnamefont {X.~Z.}\ \bibnamefont
  {Tang}}\ and\ \bibinfo {author} {\bibfnamefont {A.~H.}\ \bibnamefont
  {Boozer}},\ }\href {https://doi.org/10.1063/1.1707028} {\bibfield  {journal}
  {\bibinfo  {journal} {Physics of Plasmas}\ }\textbf {\bibinfo {volume}
  {11}},\ \bibinfo {pages} {2679} (\bibinfo {year} {2004})},\ \bibinfo {note}
  {\_eprint:
  https://pubs.aip.org/aip/pop/article-pdf/11/5/2679/19290002/2679\_1\_online.pdf}\BibitemShut
  {NoStop}%
\bibitem [{\citenamefont {Tang}(2007)}]{Tang-PRL-2007}%
  \BibitemOpen
  \bibfield  {author} {\bibinfo {author} {\bibfnamefont {X.~Z.}\ \bibnamefont
  {Tang}},\ }\href {https://doi.org/10.1103/PhysRevLett.98.175001} {\bibfield
  {journal} {\bibinfo  {journal} {Phys. Rev. Lett.}\ }\textbf {\bibinfo
  {volume} {98}},\ \bibinfo {pages} {175001} (\bibinfo {year}
  {2007})}\BibitemShut {NoStop}%
\bibitem [{\citenamefont {Riccardo}\ \emph {et~al.}(2005)\citenamefont
  {Riccardo}, \citenamefont {Loarte} \emph {et~al.}}]{riccardo2005timescale}%
  \BibitemOpen
  \bibfield  {author} {\bibinfo {author} {\bibfnamefont {V.}~\bibnamefont
  {Riccardo}}, \bibinfo {author} {\bibfnamefont {A.}~\bibnamefont {Loarte}},
  \emph {et~al.},\ }\href@noop {} {\bibfield  {journal} {\bibinfo  {journal}
  {Nuclear fusion}\ }\textbf {\bibinfo {volume} {45}},\ \bibinfo {pages} {1427}
  (\bibinfo {year} {2005})}\BibitemShut {NoStop}%
\bibitem [{\citenamefont {Stahl}\ \emph {et~al.}(2015)\citenamefont {Stahl},
  \citenamefont {Hirvijoki}, \citenamefont {Decker}, \citenamefont
  {Embr\'eus},\ and\ \citenamefont {F\"ul\"op}}]{Stahl-PRL-2015}%
  \BibitemOpen
  \bibfield  {author} {\bibinfo {author} {\bibfnamefont {A.}~\bibnamefont
  {Stahl}}, \bibinfo {author} {\bibfnamefont {E.}~\bibnamefont {Hirvijoki}},
  \bibinfo {author} {\bibfnamefont {J.}~\bibnamefont {Decker}}, \bibinfo
  {author} {\bibfnamefont {O.}~\bibnamefont {Embr\'eus}},\ and\ \bibinfo
  {author} {\bibfnamefont {T.}~\bibnamefont {F\"ul\"op}},\ }\href
  {https://doi.org/10.1103/PhysRevLett.114.115002} {\bibfield  {journal}
  {\bibinfo  {journal} {Phys. Rev. Lett.}\ }\textbf {\bibinfo {volume} {114}},\
  \bibinfo {pages} {115002} (\bibinfo {year} {2015})}\BibitemShut {NoStop}%
\end{thebibliography}%

\end{document}